\documentclass[traditabstract]{aa} 
    
\usepackage{graphicx}
 \usepackage{longtable,lscape}
\usepackage{natbib}
\usepackage{color}
\usepackage{pbox}
\usepackage{amssymb}
\usepackage[table]{xcolor}

\def\araa{ARA\&A}
\def\apj{ApJ}
\def\apjs{ApJS}
\def\mnras{MNRAS}

\newcommand{\kms}{{\rm km}\,{\rm s}^{-1}}
\newcommand{\kmss}{{\rm km}^2\,{\rm s}^{-2}}
\newcommand{\lsun}{{L}_\odot}
\newcommand{\msun}{{M}_\odot}
\newcommand{\rh}{r_{\rm h}}
\newcommand{\reff}{r_{\rm eff}}
\newcommand{\rv}{r_{\rm v}}
\newcommand{\pc}{{\rm pc}}
\newcommand{\rs}{r_0}

\newcommand{\disp}{\sigma_{\rm 1D}}
\newcommand{\dispsq}{\sigma_{\rm 1D}^2}
\newcommand{\cdisp}{\sigma_{\rm 1D}(0)}
\newcommand{\cdispsq}{\sigma_{\rm 1D}^2(0)}
\newcommand{\ml}{\Upsilon_V}

\usepackage{txfonts}
\begin{document}
   \title{The VLT-FLAMES Tarantula Survey \thanks{Based on observations collected at the European Southern Observatory under program ID 182.D-0222}}

   \subtitle{VII. A low velocity dispersion for the young massive cluster R136}

   \author{V. H\'{e}nault-Brunet\inst{1}  \and C. J. Evans\inst{2,1} \and H. Sana\inst{3}  \and M. Gieles\inst{4}  \and N. Bastian\inst{5} \and J. Ma{\'\i}z Apell{\'a}niz \inst{6} \and N. Markova\inst{7}  \and W. D. Taylor\inst{1} \and E. Bressert\inst{8,9,10} \and P.A. Crowther\inst{11} \and  J. Th. van Loon\inst{12}      }

   \institute{Scottish Universities Physics Alliance (SUPA), Institute for Astronomy, University of Edinburgh, Blackford Hill, Edinburgh, EH9 3HJ, UK ; \email{vhb@roe.ac.uk}              
              \and
              UK Astronomy Technology Centre, 
           Royal Observatory Edinburgh, 
           Blackford Hill, Edinburgh, EH9 3HJ, UK
           \and
           Astronomical Institute `Anton Pannekoek', University of Amsterdam, Postbus 94249, 1090 GE, Amsterdam, The Netherlands
           \and
           Institute of Astronomy, University of Cambridge, Madingley Road, Cambridge, CB3 0HA, UK
           \and
           Excellence Cluster Universe, Technische Universit\"{a}t M\"{u}nchen, Boltzmannstr. 2, D-85748 Garching, Germany
           \and
   Instituto de Astrof{\'\i}sica de Andaluc{\'\i}a-CSIC, Glorieta de la Astronom\'{\i}a s/n, E-18008 Granada, Spain
           \and           
           Institute of Astronomy with NAO, Bulgarian Academy of Sciences,
PO Box 136, 4700 Smoljan, Bulgaria
	\and
	School of Physics, University of Exeter, Stocker Road, Exeter EX4 4QL, UK
	\and
	European Southern Observatory, Karl-Schwarzschild-Strasse 2, D87548, Garching bei M\"{u}nchen, Germany
	\and
	Harvard-Smithsonian CfA, 60 Garden Street, Cambridge, MA 02138, USA
		\and
	Dept. of Physics \& Astronomy, Hounsfield Road, University of Sheffield, S3 7RH, UK
	\and
	Astrophysics Group, Lennard-Jones Laboratories, Keele University, Staffordshire ST5 5BG, UK
                 }

   \date{Received ; accepted }

  \abstract
{Detailed studies of resolved young massive star clusters are necessary to determine their dynamical state and evaluate the importance of gas expulsion and early cluster evolution. In an effort to gain insight into the dynamical state of the young massive cluster R136 and obtain the first measurement of its velocity dispersion, we analyse multi-epoch spectroscopic data of the inner regions of 30 Doradus in the Large Magellanic Cloud (LMC) obtained as part of the VLT-FLAMES Tarantula Survey. Following a quantitative assessment of the variability, we use the radial velocities of non-variable sources to place an upper limit of 6\,$\kms$ on the line-of-sight velocity dispersion of stars within a projected distance of 5\,pc from the centre of the cluster. After accounting for the contributions of undetected binaries and measurement errors through Monte Carlo simulations, we conclude that the true velocity dispersion is likely between 4 and 5\,$\kms$ given a range of standard assumptions about the binary distribution. This result is consistent with what is expected if the cluster is in virial equilibrium, suggesting that gas expulsion has not altered its dynamics. We find that the velocity dispersion would be $\sim$25\,$\kms$ if binaries were not identified and rejected, confirming the importance of the multi-epoch strategy and the risk of interpreting velocity dispersion measurements of unresolved extragalactic young massive clusters.} 

   \keywords{ binaries: spectroscopic -- galaxies: star clusters: individual (R136) -- Magellanic Clouds -- stars: early-type -- stars: kinematics and dynamics  }

\authorrunning{V. H\'{e}nault-Brunet et al.}
\titlerunning{VFTS VII. A low velocity dispersion for R136}

   \maketitle
%

\section{Introduction}

The expression ``infant mortality" of star clusters was initially coined by \citet{lada2003} to describe the discrepancy between the number of observed open clusters and the number of embedded clusters. These authors argued that there are about ten times fewer open clusters than expected if all embedded clusters evolve into open clusters. The rapid removal of gas left-over from star formation was suggested to explain the apparent disruption of such a large fraction of clusters \citep[e.g.][]{geyer, kroupaboily, BG2006}. The importance of the infant mortality scenario however depends on the definition adopted for embedded clusters \citep{bressert2010, bastian2011}, with more conservative criteria requiring less than 50\% of clusters to be destroyed to match the observed number of open clusters. But no matter which definition is adopted, the question of whether or not gas expulsion plays a significant role in the early evolution/disruption of star clusters still needs to be addressed.

Star clusters have been observed to expand in their first 100\,Myr \citep{2003MNRAS.338...85M, bastian2008, 2010ARA&A..48..431P}, but this expansion is not direct evidence for the importance of gas expulsion. There are two ways for clusters to expand as a response to mass loss \citep[e.g.][]{hills1980}: (i) expansion following impulsive mass loss, e.g. change of potential due to removal of mass faster than the crossing time of the cluster, leaving the cluster in a super-virial state for a few crossing times, or (ii) adiabatic expansion, e.g. driven by stellar evolution on a slow timescale compared to the crossing time of the cluster \citep[$\sim$10\,Myr for young massive clusters, e.g. ][]{2010ARA&A..48..431P}, in which case the cluster remains in virial equilibrium. Thus, the best way to evaluate the importance of rapid gas expulsion (case i) and the implications for the formation and early evolution of star clusters is to determine the dynamical state of young clusters. In particular, it is important to verify if clusters are in virial equilibrium from a young age.

Attempts to determine the dynamical state of young clusters have been made by comparing dynamical masses (obtained through measuring the velocity dispersion and size of a cluster) and photometric masses (estimated from the age and integrated luminosity). For several unresolved extragalactic star clusters with ages of less than $\sim$10\,Myr, the dynamical mass has been found to be up to ten times larger than the photometric mass \citep[e.g.][]{bastianetal:2006}. This led to the suggestion that these clusters might be super-virial and expanding following gas expulsion \citep{goodwinbastian:2006}. However, \citet{gieles2010} showed that the increase in the measured line-of-sight velocity dispersion in these young clusters could be produced by the orbital motions of massive binaries. Massive stars indeed dominate the light of young massive clusters and their binary fraction is high \citep[e.g.][Sana et al., submitted - hereafter Paper VIII]{mason2009, barba2010, sanaevans}. Therefore, the role of gas expulsion cannot be investigated through observations of unresolved extragalactic clusters.

Detailed dynamical studies of individual resolved clusters (i.e. in the Galaxy, Small Magellanic Cloud - SMC - or Large Magellanic Cloud - LMC) are necessary to make progress. In the case of radial velocity (RV) surveys, multi-epoch observations are needed to detect binaries, which can then be removed and allow a cleaner estimate of the velocity dispersion (unless the orbital solution is known, in which case the centre-of-mass velocity can be included in the dispersion calculation). The young massive cluster R136 \citep[$M\sim10^{5}$~$\msun$,][]{2009ApJ...707.1347A}, in the 30~Doradus region of the LMC, is an ideal target to test the impact of gas expulsion given its young age of less than 2\,Myr \citep{koterheap, massey1998, Crowther2010}.

The identification of binaries was central to the study of stellar dynamics in 30 Doradus with the Gemini Multi-Object Spectrograph by \citet{Bosch:2009}, who found that the velocity dispersion of NGC\,2070 went from $\sim$30\,$\kms$ to 8.3\,$\kms$ after correcting for orbital motions, confirming their prediction \citep{bosch2001} that the high velocity dispersion of NGC 2070 could be due to undetected binaries. In order to perform a similar study for the dense surroundings of R136, a different observational approach was required because this region is too crowded for fibre spectroscopy. A key component of the VLT-FLAMES Tarantula Survey \citep[VFTS;][hereafter Paper I]{evans:2011} is multi-epoch spectroscopy in the inner part of 30 Dor with the FLAMES--ARGUS integral-field unit (IFU), which we use here to obtain an estimate of the velocity dispersion of `single' stars in R136. The central velocity dispersion is an important proxy for the dynamical state of the cluster and is required to estimate, for example, the central potential and relaxation time-scale, both of which are required for detailed numerical ($N$-body) calculations of R136-like objects \citep[e.g.][]{PZ:1999, gieles_mr:2010}. As a result of the same study, the identification of massive binaries could provide important clues to the star-formation process and the subsequent dynamical evolution of R136, in which binaries can be formed, ejected and destroyed. Our measurements will also serve as useful empirical input for the modeling of stellar interactions in dense clusters, such as the recent studies of \citet{fujii} and \citet{banerjee}.

We report here on a velocity dispersion estimate for the young massive cluster R136. We describe the FLAMES--ARGUS IFU data in Sect.~\ref{argus_data}. In Sect.~\ref{variability}, we present our RV and variability analysis of the stars observed with ARGUS. In Sect.~\ref{class}, we discuss the spectral classification of the non-variable ARGUS sources.  In Sect.~\ref{medusa}, we briefly introduce the VFTS Medusa data complementing the ARGUS data. We calculate the velocity dispersion from the stars showing no variability and also estimate the contribution of errors and undetected binaries in Sect.~\ref{vdisp}. The implications of the measured velocity dispersion for the dynamical state of R136 are discussed in Sect.~\ref{discuss}. Finally, we present our conclusions in Sect.~\ref{conc}.

\section{ARGUS data \label{argus_data}}

The main data used in this paper consist of five ARGUS IFU pointings in the central arcminute of 30~Dor (see Figure 2 of Paper I), for which at least five epochs were obtained. This central region is relatively gas free, so the nebular contamination in the spectra is not as important as in other regions of 30~Dor. Spectra of 41 different sources, corrected to the heliocentric frame, were extracted from the data cubes. To probe the stellar dynamics of R136, the LR02 setting of the Giraffe spectrograph was used ($\lambda$=3960--4570 \AA, $\Delta \lambda=0.40$ \AA, R$\equiv \lambda/\Delta\lambda \sim$10\,500). This setting gives access to several stellar absorption lines suitable for RV  analysis (see section \ref{variability}) at the typical signal-to-noise ratio (S/N) of $\sim$85 of our single-epoch spectra. To optimize the detection of binaries, the first two epochs were observed without time restrictions, and the third and fourth were observed with a minimum interval of 28 days between both the second and third, and third and fourth epochs.  The fifth epoch was observed at least one year after the first epoch. Some observations were repeated due to changes in conditions and other operational issues, providing additional epochs for three of the five pointings. Note that expanding the time baseline of the observations enhances the chances of detecting long-period binaries, but these binaries are the ones that have a smaller impact on the velocity dispersion enhancement. Further details on the ARGUS data, the reduction and the extraction procedure can be found in Paper I.

For the analysis presented in this paper, individual exposures were considered to belong to the same epoch if their start time was separated by less than one hour. The multiple exposures composing a single epoch were then averaged using the errors (propagated throughout the reduction process) as weights and performing a 5\,$\sigma$ clip around the median to remove remaining cosmic features. The spectra from individual exposures had already been normalised as part of the extraction procedure. The resulting epochs, their modified Julian date, and the corresponding ARGUS pointings and exposures are listed in Table \ref{epochs}.

An effort was made to extract spectra preferably for sources that appeared single by comparing the ARGUS data cubes with an archival HST Wide-Field Camera Three (WFC3) F555W image \citep{deMarchi2011} and identifying matching sources (see Paper I). Out of 41 ARGUS sources, 23 are dominated by one bright source from the WFC3 image. For these 23 ARGUS sources, only stars at least a factor of ten fainter (in data counts) are visible in the WFC3 image within the region covered by the ARGUS spatial pixels from which the spectrum was extracted. Even if those fainter stars were contaminating the ARGUS spectrum, it is doubtful that they would contribute significantly to the helium absorption lines used for our RV analysis (section \ref{var_crit}). The ARGUS spatial elements for a further 11 extracted sources are dominated by one bright object from the WFC3 image, but could suffer from more significant contamination from nearby stars (typically at a level of $\sim20\%$). The remaining seven ARGUS sources appear multiple in the WFC3 F555W image, with two or more densely-packed bright stars contributing at a comparable level to the flux in the region of the ARGUS source. These seven ARGUS sources were retained  because they could still prove useful to our analysis (see below).

Note that even the spectrum of apparently single ARGUS sources (based on the WFC3 image) could contain contributions from multiple stars. The inner part of R136 is densely populated with stars. Two stars could be several thousand AU apart and, given the distance to the LMC, still appear as a single source in the WFC3 image. We would not expect to detect the motion of unresolved binaries with such large separations given our spectral resolution and time coverage. These considerations are however not real concerns for our study because we can identify shorter period binaries from RV variations (section \ref{var_crit}) and estimate the residual contribution of undetected binaries to the velocity dispersion using Monte Carlo simulations (section \ref{effect_binaries}). Stars showing double/asymmetric line profiles or inconsistent absolute RVs between different lines can also be flagged as multiple (either true binaries or a chance alignment of stars along the line of sight). On the other hand, we can still use the sources which appear multiple in the WFC3 image if they show none of the above spectroscopic signs of binarity/multiplicity. In that case, one star could be dominating the spectrum, or several stars could be contributing without showing any apparent RV difference, and the source would still be valid to study the dynamics of the cluster.

\section{Radial velocity and variability analysis \label{variability}}

\subsection{Zero-point errors}

\begin{figure}[!htb]
   \centering
\includegraphics[width=8cm]{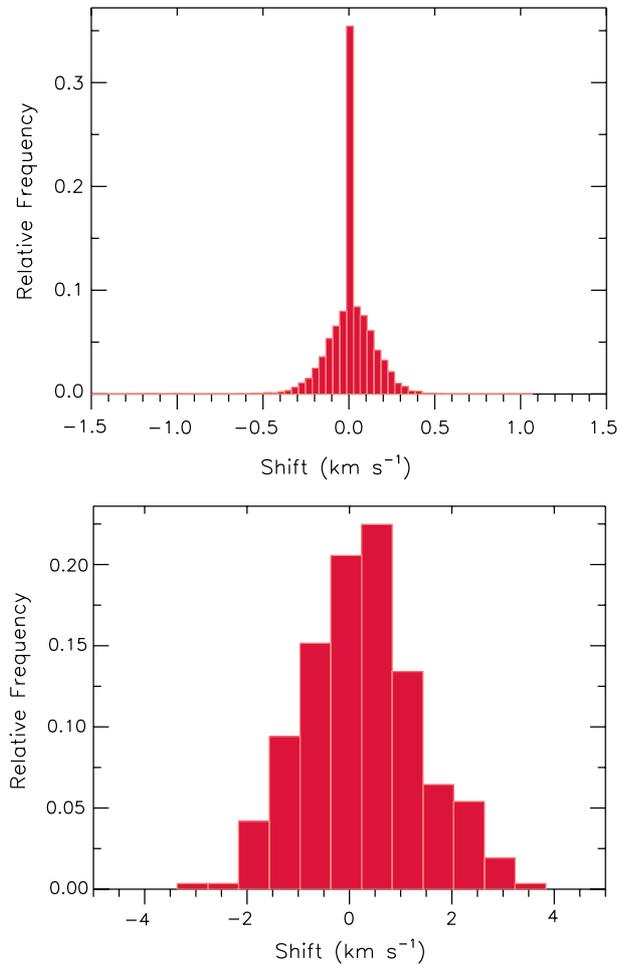}
      \caption{Distribution of velocity shifts from the cross-correlation of ARGUS calibration arc spectra (top) and from the cross-correlation of nebular lines (bottom) between different epochs/exposures.}
         \label{fig:zeropoint}
   \end{figure}

To check that zero-point errors do not affect our multi-epoch RV measurements significantly, we first cross-correlated the ARGUS calibration arc spectra of each epoch/exposure with the arc spectrum of the corresponding fibre from the first exposure of the first epoch. We then determined the zero-point velocity shifts from the peak of the resulting cross-correlation functions. The corresponding distribution of zero-point errors for over 20\,000 such measurements is shown in Fig.~\ref{fig:zeropoint} (top panel). The vast majority of the measurements lie between $-0.5$ and 0.5~$\kms$ \ with a peak at $\sim$0~$\kms$, suggesting that the instrument wavelength calibration is remarkably stable and does not introduce spurious variations of stellar radial velocity between the different epochs. To investigate possible issues with the telescope and/or instrument not accounted for in the arc spectra, we also performed a similar analysis on the nebular lines (H${\gamma}$ and H${\delta}$) from selected spaxels with minimum stellar contamination. The resulting distribution of velocity shifts for over 500 measurements peaks near 0~$\kms$ and has a small standard deviation of $\sim$1~$\kms$ (Fig.~\ref{fig:zeropoint}, bottom panel), which again suggests that there are no significant systematic shifts between epochs.

\subsection{Variability criteria \label{var_crit}}

In order to exclude the stars exhibiting RV variations from our calculation of the velocity dispersion, we implemented the following quantitative assessment of the variability.

\begin{figure}[!t]
   \centering
\includegraphics[width=9cm]{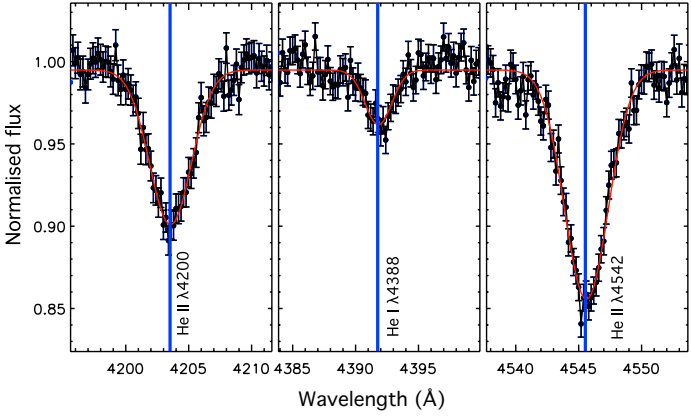}
      \caption{Example of a simultaneous Gaussian fit (i.e. same RV for all lines) to the \ion{He}{ii}~$\lambda$4200, \ion{He}{i}~$\lambda$4388, and \ion{He}{ii}~$\lambda$4542 absorption lines for an individual epoch of the ARGUS source VFTS~1026 (O2-4.5 + mid/late~O, see section~\ref{class}).}
         \label{Gaussfit}
   \end{figure}

For each star and epoch, we started by fitting Gaussian profiles to individual stellar absorption lines (\ion{He}{ii}~$\lambda$4200, \ion{He}{ii}~$\lambda$4542, \ion{He}{i}+\ion{He}{ii}~$\lambda$4026, \ion{He}{i}~$\lambda$4143, \ion{He}{i}~$\lambda$4388 and \ion{He}{i}~$\lambda$4471, or a subset when some of these lines were too weak or the S/N too low). In principle, the profiles of the He lines are not Gaussian, particularly in the wings. However, given the moderate S/N of our spectra and the noise in the line wings, the use of Gaussians provided good fits to the line profiles (see Fig.~\ref{Gaussfit}) and did not affect the results. We used the MPFIT {\sc idl} least-squares fitting routine \citep{mpfit} and Gaussian profiles defined by their central radial velocity, width, depth, and continuum value. An example fit is shown in Fig.~\ref{Gaussfit} for the ARGUS source VFTS~1026. The resulting 1$\sigma$ error on the measured RV in this case is $\pm$2.1~$\kms$, illustrating the precision that can be achieved for a single epoch with a good quality spectrum.  To check if errors in the normalisation of the continuum could
influence the RV measurements, we also performed the fits allowing for a linear component instead of a constant continuum, but this did not affect the results. We chose the Gaussian fitting approach (as opposed to cross-correlation for example) because it can provide reliable error estimates on the velocities and directly yields absolute RVs, which we need to compute the velocity dispersion if different lines are used for different stars (see section \ref{abs}). This method is also well suited to our data set, for which the quality of spectra varies significantly between stars and even from one epoch to the next.

The lines listed above are characteristic of the O-type stars composing the vast majority of our sample (see section~\ref{class} and Table~\ref{rv_table}). They were the only lines generally strong enough to perform satisfying RV measurements on the spectra of individual epochs. Metallic lines were not considered because they were generally too weak, and Balmer lines were ignored because their profiles and apparent RVs can be affected by stellar winds or strong/variable nebular contamination. When one of the fitted lines appeared significantly contaminated by nebular emission it was also rejected, unless the nebular component could be clearly identified and the fit could be performed on the wings of the line only. The most affected line was \ion{He}{i}~$\lambda$4471, but it was also occasionally a problem for the other \ion{He}{i} lines.

From close inspection of all the fitted profiles, we first identified the stars showing double-lined spectroscopic binary (SB2) profiles. Four of the 41 ARGUS sources show SB2 profiles (VFTS 1016, 1019, 1031 and 1033), and none of them had a primary component with a constant RV throughout all epochs (based on the criteria below), suggesting that they are genuine multiple systems and not just the result of the alignment of two stars with different RVs along the line of sight. The epochs at which SB2 profiles are observed in these sources are indicated in Table \ref{var}.

We considered a star as a radial velocity variable if the series of RV measurements of any of the fitted lines of that star contained at least one strongly deviant point, which we defined as

\begin{equation}
\left | {\frac{RV_{i} - \mu}{\sigma_{i}}} \right | > 4,
\label{4sig}
\end{equation}

\noindent{where $RV_{i}$ and $\sigma_{i}$ are the radial velocity and its 1\,$\sigma$ error at epoch $i$, and $\mu$ is the weighted mean RV over all the epochs. If the above condition was fulfilled for at least one individual line, then the null hypothesis of constant RV was rejected.

 For each star and each line, we also computed the value of $\chi^2$ assuming a constant RV, i.e.

\begin{equation}
\chi^2 = \sum_{i} \frac{(RV_{i} - \mu)^2}{\sigma_{i}^2},
\label{chi2}
\end{equation}

\noindent{and rejected the constant RV hypothesis when the goodness of fit of a constant RV model to the data was poor, which we defined as

\begin{equation}
1 - P(\chi^2, \nu) < 10^{-4},
\label{pnu}
\end{equation}

\noindent{where $P(\chi^2,\nu)$ is the probability that, in a $\chi^2$ distribution with $\nu$ degrees of freedom ($\nu=$\# of epochs $-$ 1), the value of $\chi^2$ is less than or equal to the value computed in Eq. \ref{chi2}. The thresholds adopted in Eqs. \ref{4sig} and \ref{pnu} were chosen so that the probability of a false variability detection in our sample (given the sampling and accuracy of our measurements) remained negligible. In their analysis of the multiplicity properties of the O-type stars in VFTS, Sana et al. (Paper VIII) adopted slightly different variability criteria, but mention that their results are generally equivalent to those of a variability test based on the goodness of fit of a constant RV model like that of Eqs. \ref{chi2} and \ref{pnu}.

As a further check, we investigated line profile variations (LPV) by computing Time Variance Spectra \citep[TVS;][]{Fullerton:1996}. The only difference with the method of \citet{Fullerton:1996} is that we used the known error bars at each pixel to define a wavelength-dependent threshold instead of using a flat threshold based on continuum noise. This has the advantage of taking into account the varying S/N as a function of wavelength. The TVS at wavelength $\lambda$ is given by

\begin{equation}
TVS (\lambda) = \frac{1}{N-1} \frac{N}{\sum_{i}1/ \sigma_{i}^2(\lambda)} \sum_{i} \frac{(f_{i}(\lambda) - \langle f(\lambda) \rangle )^2}{\sigma_{i}^2(\lambda)},
\label{tvseq}
\end{equation}

\noindent{where $N$ is the number of spectra in the time series, $f_{i}(\lambda)$ and $\sigma_{i}(\lambda)$ are the flux and 1\,$\sigma$ error at wavelength $\lambda$ for epoch $i$, and $\langle f(\lambda) \rangle$ is the weighted average flux of all epochs at wavelength $\lambda$. The threshold for variability at a confidence level of 99\% is

\begin{equation}
TVS (\lambda) > \frac{N}{\sum_{i}1/ \sigma_{i}^2(\lambda)} \frac{P_{\chi^2}(0.01, N-1)}{N- 1},
\label{thresh}
\end{equation}

\noindent{where $P_{\chi^2}(0.01, N-1)$ is the cutoff value in a $\chi^2$ distribution with $N-1$ degrees of freedom such that the probability that a random  variable is greater than this cutoff value is equal to 0.01. The cases for which significant variability is inferred from the TVS are indicated in Table \ref{var}. These results generally confirm those obtained from the other variability tests, and in two cases help to establish significant variability in emission-line stars with no or only weak absorption lines. Fig. \ref{TVS} shows examples of TVS for two stars of our ARGUS sample, one revealing no variability and the other showing significant LPV. Note that the TVS often shows variability in nebular emission lines due to changes in conditions affecting the sky subtraction, but we ignore these spectral regions when assessing the variability of a star. The quantity TVS$^{1/2}$ is plotted in Fig. \ref{TVS} because it scales linearly with the size of the spectral flux deviations and gives a direct estimate of the amplitude of the variations.

If none of the three tests above (Eq. \ref{4sig}, Eq. \ref{pnu}, TVS) revealed variability in individual lines, we performed simultaneous Gaussian fits of \ion{He}{ii}~$\lambda$4200, \ion{He}{ii}~$\lambda$4542 and \ion{He}{i}~$\lambda$4388 (or a subset of these lines when one of them was too weak or simply not present) by forcing their central RV to be the same (see Fig.~\ref{Gaussfit} for an example). These three lines were adopted because they give reliable absolute RVs (see section \ref{abs}). We can therefore fit them together to obtain more precise RV measurements. The series of RV measurements resulting from these simultaneous fits were then tested for variability using Eqs. \ref{4sig} and \ref{pnu} again (see results in Table \ref{var}). Stars still showing no sign of variability were then considered as suitable to study the dynamics of the cluster. The RVs of the ARGUS sources for all individual epochs are presented in Table~\ref{RV_ind}.

Out of 41 ARGUS sources, 16 are not detected as variable, 17 are variable (four SB2, 11 SB1, two emission-line stars with variability determined from the TVS), seven have a too low S/N for a meaningful variability analysis, and another one is an emission-line star with no suitable absorption line for RV analysis and for which no significant variability was detected from the TVS.

   \begin{figure}[!t]
   \centering
\includegraphics[width=9cm]{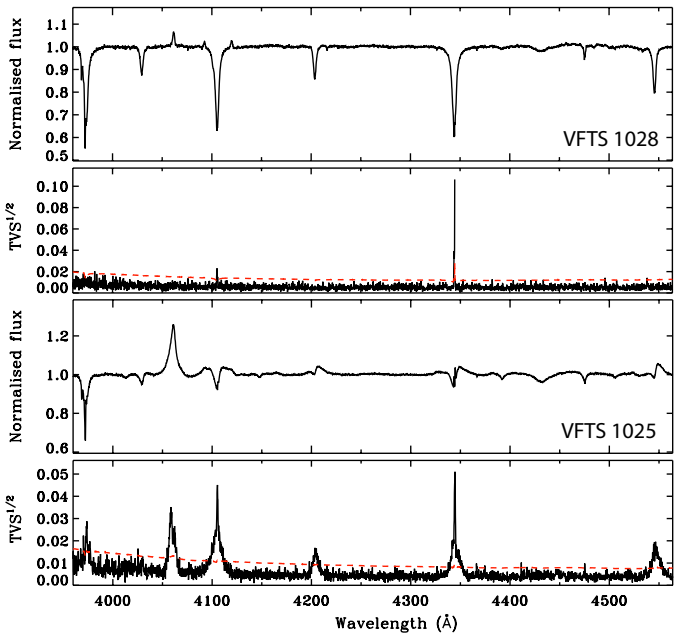}
      \caption{Weighted mean spectrum (first and third panels from the top) and Temporal Variance Spectrum (TVS$^{1/2}$; second and fourth panels from the top) for the ARGUS sources VFTS~1028 and VFTS~1025. The red dashed curves indicate the 99\% confidence level for variability. Significant variability is only seen in the nebular emission lines of VFTS~1028, but it is detected in several stellar lines in VFTS~1025.}
         \label{TVS}
   \end{figure}

\subsection{Absolute radial velocities \label{abs}}

We want to use as many lines as possible to increase the precision of RV measurements, but at the same time we need to make sure that the selected lines give accurate results. In particular, all the lines fitted simultaneously to measure the RV of a given star should give consistent results when fitted individually, and the selected lines should provide consistent absolute RVs if different subsets of lines are used for different stars (e.g. \ion{He}{i} lines for late O-type stars and \ion{He}{ii} lines for early O-type stars).

Our choice of suitable lines for absolute RV measurements (\ion{He}{i}~$\lambda$4388, \ion{He}{ii}~$\lambda$4200, and \ion{He}{ii}~$\lambda$4542) is supported by the RV analysis performed in Paper VIII on the large sample of O-type stars observed with Medusa (see section~\ref {medusa}). The final RVs adopted for the non-variable ARGUS stars are obtained from simultaneous fits of these three lines or a subset of them.

In stars with strong winds, wind infilling of photospheric lines can modify the RV measured by Gaussian fitting for lines like \ion{He}{ii}~$\lambda$4200 and \ion{He}{ii}~$\lambda$4542 and also result in RV shifts between these lines. Comparison of Gaussian fitting measurements and values determined from CMFGEN models shows good consistency for O-type dwarfs and giants, but small shifts of a few tens of $\kms$ \ for supergiants (P. Crowther, private communication). In our calculation of the velocity dispersion (section~\ref{upper}), we therefore payed particular attention to the stars that are identified as possible supergiants (section~\ref{class}).

\section{Spectral classification of ARGUS non-variable sources \label{class}}

   \begin{figure*}[!t]
   \centering
\includegraphics[width=16cm]{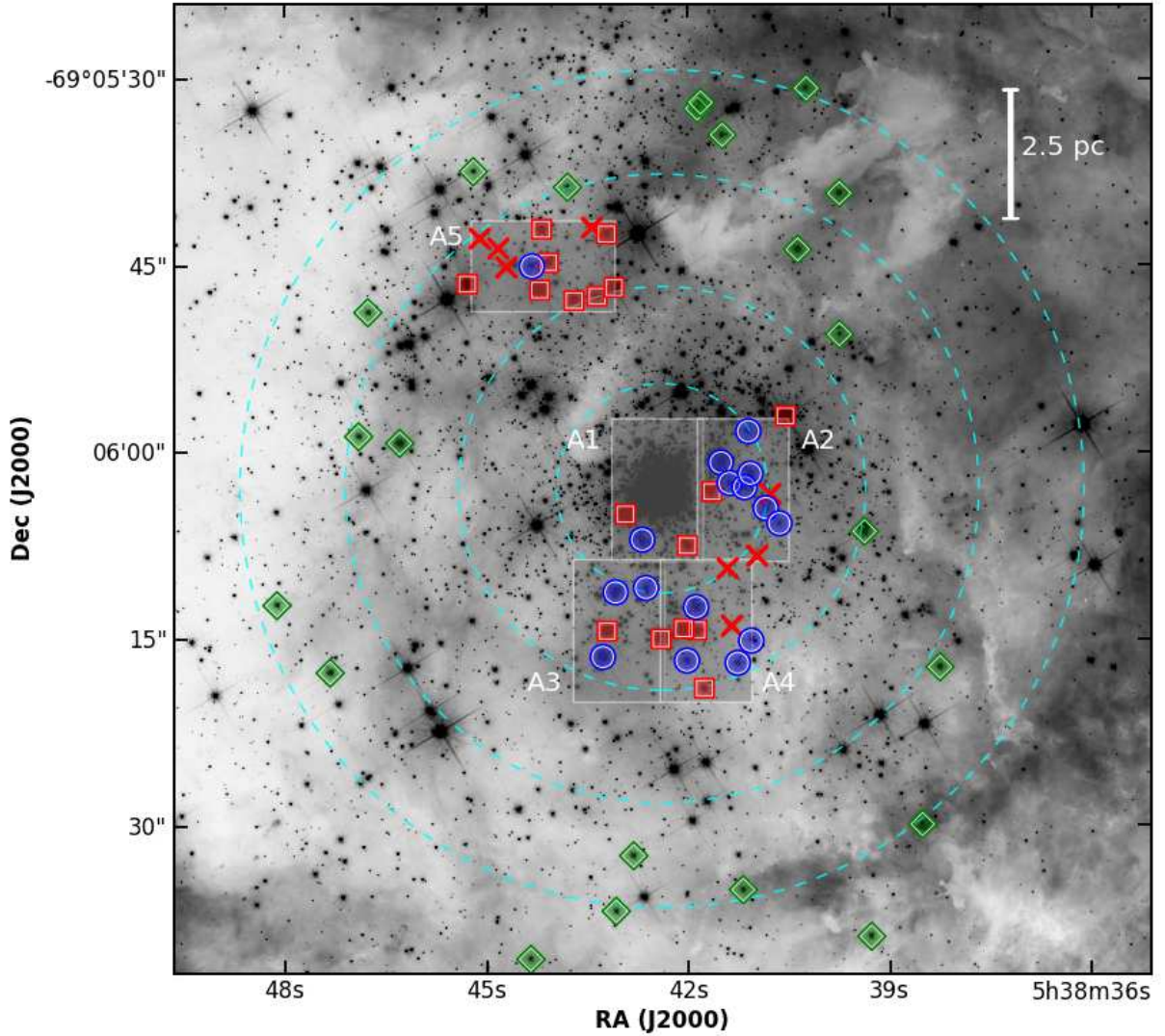}
      \caption{Distribution of ARGUS and Medusa sources used in this work overlaid on an F555W {\em HST}-WFC3 image. ARGUS stars in which no variability was detected are shown with blue circles, ARGUS variable stars are represented by red squares, and ARGUS stars with too low S/N for RV analysis or no suitable absorption lines are indicated by red crosses. The ARGUS IFU pointings (A1 to A5) are also shown as grey transparent rectangles. Medusa non-variable stars added to the sample are represented by green diamonds. The dashed circles indicate projected radial distances of 2, 4, 6, and 8\,pc from the centre of R136.}
         \label{argus}
   \end{figure*}

\begin{table*}[!t]

\centering                 
\caption{RVs of ARGUS and MEDUSA sources showing no significant variability.}

\footnotesize{            
\label{rv_table}      
\begin{tabular*}{18cm}{ l l l c c c  c  l  }       
\hline\hline                 
 \multicolumn{2}{l}{ID} & Data  & $\alpha$ & $\delta$  & r$_{\rm d}$ & RV & Spectral type\\     
\cline{1-2}  VFTS & Aliases$^{*}$ & & (2000) & (2000)  & [pc]  & [km~s$^{-1}$] &  \\
\hline
\rowcolor[gray]{.88} 1024 & S99-147, R136-226, MH-623 & ARGUS& 05 38 42.685 & $-$69 06 07.03 & 1.1    &263.7$\pm$2.3 & O8.5 III/V \\
1014 & S99-56, R136-29, MH-203, P93-863 & ARGUS& 05 38 41.515 & $-$69 06 00.83  & 1.2    &266.5$\pm$0.7 & O2-4.5+mid/late O   \\
&    & &  &  &  &    &O3 V [MH98] \\
\rowcolor[gray]{.88}1012& S99-249, R136-151, MH-178 & ARGUS& 05 38 41.386 & $-$69 06 02.49 & 1.3   &264.6$\pm$2.8 & O9 III/V \\
1009& S99-165, R136-88, MH-141 & ARGUS& 05 38 41.163 & $-$69 06 02.83  &1.6  &274.2$\pm$1.4 & O6.5 III/V \\
\rowcolor[gray]{.88}1007 & S99-95, R136-60, MH-129, P93-827& ARGUS& 05 38 41.077 & $-$69 06 01.74 & 1.7  &264.1$\pm$1.3 & O6.5 III/V \\
1004&  S99-193, R136-126, MH-95 & ARGUS& 05 38 40.848 & $-$69 06 04.51& 2.0    & 274.4$\pm$2.0 & O9.5 III/V\\
\rowcolor[gray]{.88}1008& S99-163, R136-96, MH-134 & ARGUS& 05 38 41.108& $-$69 05 58.33  & 2.0   &277.3$\pm$1.0 & ON6.5 I/II \\
1023& S99-142, R136-83, MH-591 & ARGUS& 05 38 42.631 & $-$69 06 10.91 & 2.0   &266.4$\pm$2.0 & O8 III/V \\
&  &  & &  &  &    & O6 V [MH98]\\

\rowcolor[gray]{.88}1026& (S99-191), R136-41, MH-716, Mk35N& ARGUS& 05 38 43.083 & $-$69 06 11.26 & 2.2   &265.0$\pm$0.9& O2-4.5+mid/late O \\
\rowcolor[gray]{.88} & &  &  &      &       & & O3 III(f*) [MH98]\\
\rowcolor[gray]{.88}  & &  &     &   &       & & O8: V [WB97]\\

1002 &S99-312, R136-194, MH-77& ARGUS& 05 38 40.646 & $-$69 06 05.73  & 2.4   & 272.7$\pm$5.6 & O9.5 III/V \\
\rowcolor[gray]{.88}1018&  S99-88, R136-37, MH-290, P93-900 & ARGUS& 05 38 41.887 & $-$69 06 12.45  & 2.4 &258.4$\pm$0.6& O2-4.5+mid/late O\\
\rowcolor[gray]{.88} &    &  &  &     &   & & O3 III(f*)  [MH98]\\
1020& S99-178, R136-101, MH-314 & ARGUS& 05 38 42.023 & $-$69 06 16.75  & 3.4   &267.4$\pm$0.9& O3-4  \\
\rowcolor[gray]{.88}1006& S99-257  & ARGUS& 05 38 41.066 & $-$69 06 15.16  & 3.4   &264.2$\pm$3.9& O6.5 III/V \\
1028&  S99-37, R136-23, P93-1036, Mk35S& ARGUS& 05 38 43.274 &$-$69 06 16.45  & 3.5   &271.9$\pm$0.5& O3.5-4.5 \\ 
 &    &  &   &     & &   & O3 III(f*)  [MH98]\\
  &    &  &     & & &   & O4-5 V:  [WB97]\\
\rowcolor[gray]{.88}1010&  S99-189, R136-103, MH-156 & ARGUS& 05 38 41.268 & $-$69 06 16.94 & 3.7 &283.1$\pm$2.2 & O7 III/V \\
468& S99-86, MH-17, P93-706, Mk36& Medusa & 05 38 39.38$\phantom{~~}$ &  $-$69 06 06.39  & 4.0   & 272.0$\pm$0.8& O2 V((f*)) + OB  \\ 
\rowcolor[gray]{.88}477& S99-455& Medusa & 05 38 39.75$\phantom{~~}$ &  $-$69 05 50.55  & 4.5   & 265.3$\pm$3.5&O((n)) \\ 
1035& S99-169, R136-109 & ARGUS& 05 38 44.321& $-$69 05 45.05  & 5.0  &268.5$\pm$2.2& O8.5 I/II  \\
 &    &  &   &      & &   & O8 V  [MH98]\\
\rowcolor[gray]{.88}601& S99-91, MH-986, P93-1317, Mk14N & Medusa & 05 38 46.29$\phantom{~~}$ &  $-$69 05 59.25  & 5.1   & 266.8$\pm$0.6 &O5-6 V((n))z   \\ 
\rowcolor[gray]{.88}  & &  &     &   &       & & O5 V((f))  [MH98] \\
\rowcolor[gray]{.88}  & &  &     &   &       & & O4 V  [WB97] \\
484& S99-124 & Medusa & 05 38 40.37$\phantom{~~}$ &  $-$69 05 43.72  & 5.3    & 284.2$\pm$1.0 & O6-7 V((n))  \\
\rowcolor[gray]{.88}611 &S99-270 & Medusa & 05 38 46.90$\phantom{~~}$ &  $-$69 05 58.71 & 5.9   &  265.4$\pm$2.2&O8 V(n)   \\
554& S99-343 & Medusa & 05 38 43.79$\phantom{~~}$ & $-$69 05 38.70 & 6.1   & 275.6$\pm$1.4& O9.7 V  \\
\rowcolor[gray]{.88}446& S99-194 & Medusa & 05 38 38.26$\phantom{~~}$ &  $-$69 06 17.29  & 6.4   & 258.9$\pm$5.0 & O Vnn((f))  \\
607& S99-294 & Medusa & 05 38 46.76$\phantom{~~}$ &  $-$69 05 48.75  & 6.6     & 257.8$\pm$1.0 & O9.7 III  \\
\rowcolor[gray]{.88}476& S99-206 & Medusa & 05 38 39.75$\phantom{~~}$ &  $-$69 05 39.21  & 6.7   & 270.1$\pm$2.2 & O((n))  \\ 
505& S99-265& Medusa & 05 38 41.49$\phantom{~~}$  & $-$69 05 34.52  & 7.0   &265.5$\pm$3.4& O9.5 V-III  \\
\rowcolor[gray]{.88}536& S99-295 & Medusa & 05 38 42.82$\phantom{~~}$ &  $-$69 06 32.43 & 7.2  & 248.4$\pm$1.4 & O6 Vz  \\
582& S99-414 & Medusa & 05 38 45.19$\phantom{~~}$ &  $-$69 05 37.43  & 7.2   & 270.2$\pm$2.3&O9.5 V((n))  \\
\rowcolor[gray]{.88}615 &S99-218 & Medusa & 05 38 47.33$\phantom{~~}$ &  $-$69 06 17.70  & 7.3  & 256.4$\pm$7.6&O9.5 IIInn  \\
515& S99-434 & Medusa & 05 38 41.86$\phantom{~~}$ &  $-$69 05 32.40  & 7.4   & 267.4$\pm$2.0 &O6-9p  \\ \rowcolor[gray]{.88}513&S99-266 & Medusa & 05 38 41.81$\phantom{~~}$ &  $-$69 05 31.91  & 7.6  & 266.3$\pm$1.3  & O6-7 III-II(f)  \\ 
622 & S99-333  & Medusa & 05 38 48.12$\phantom{~~}$ &  $-$69 06 12.27  & 7.8   & 272.0$\pm$1.6&O9.7 III  \\ 
\rowcolor[gray]{.88}498& S99-347& Medusa & 05 38 41.19$\phantom{~~}$ &  $-$69 06 35.17 & 8.0   & 265.9$\pm$7.2 & O9.5 V  \\
451& S99-346  & Medusa & 05 38 38.52$\phantom{~~}$ &  $-$69 06 29.92  & 8.3   & 277.3$\pm$ 10.8 &O9: III:(n)  \\
\rowcolor[gray]{.88}483& S99-309 & Medusa & 05 38 40.24$\phantom{~~}$ &  $-$69 05 30.78  & 8.3    & 275.4$\pm$2.7&O9 V  \\
540& S99-372 & Medusa & 05 38 43.08$\phantom{~~}$ &  $-$69 06 36.88  & 8.3   & 242.7$\pm$1.1& B0 V  \\
\rowcolor[gray]{.88}560& S99-350, P93-1139 & Medusa & 05 38 44.35$\phantom{~~}$ & $-$69 06 40.73  & 9.5   & 259.7$\pm$0.9& O9.5 V  \\
465 &S99-365, P93-700 & Medusa & 05 38 39.28$\phantom{~~}$ &  $-$69 06 38.93 & 9.6   & 255.8$\pm$4.4 & O5: Vn  \\
\hline

\end{tabular*}
}
\tablebib{MH = \citet{massey1998}, WB97 = \citet{walborn1997}. $^{*}$Aliases identification numbers are from \citealt{selman1999} (S99-), \citealt{hunter1997} (R136-), \citealt{malheap} (MH-), \citealt{parker1993} (P93-) and \citealt{Mk85} (Mk).}
\tablefoot{Spectral types were determined in this work (see section~\ref{class}) for the ARGUS sources and by Walborn et al. (in preparation) for the Medusa sources, unless otherwise indicated. The sources are sorted by increasing projected radial distance (r$_{\rm d}$) from R136-a1 ($\alpha$ = 5$^{\rm h}$38$^{\rm m}$42$^{\rm s}$. 39, $\delta$ = $-$69$\degr$06'02.''91, J2000).}

\end{table*}

To get a general idea of the spectral content of our ARGUS sample, we classified the non-variable and presumably single stars. We did not attempt the complex task of classifying the binaries/variable stars, partly because the limited wavelength coverage of the ARGUS spectra makes it even more difficult. 

To assign spectral types (SPT), we visually inspected the ARGUS spectra degraded to an effective resolving power of 4\,000 
and, following the premises of \citet{sota}, we performed 
a morphological classification. In particular, for stars at intermediate 
and late subtypes, we used the eye-estimated line ratios of \ion{He}{i+ii}~$\lambda$4026 to \ion{He}{ii}~$\lambda$4200, \ion{He}{i}~$\lambda$4471 to \ion{He}{ii}~$\lambda$4542, \ion{He}{i}~$\lambda$4388 to \ion{He}{ii}~$\lambda$4542, \ion{He}{i}~$\lambda$4143 to \ion{He}{ii}~$\lambda$4200 and  \ion{Si}{iii}~$\lambda$4552 to \ion{He}{ii}~$\lambda$4542 in order to assign the spectral subtype. For the hottest stars, on the other hand, we were not able to exploit the primary criteria based on the Nitrogen ionization equilibrium because our ARGUS spectra do not cover wavelengths beyond $\sim$4570 \AA. We thus concentrated on criteria related to the initial appearance of certain 
\ion{He}{i} lines (such as \ion{He}{i}~$\lambda$4471, \ion{He}{i}~$\lambda$4143, and \ion{He}{i}~$\lambda$4388), and the presence and strength of \ion{N}{iv}~$\lambda$4058 and the \ion{Si}{iv} doublet in emission. The morphological classification of our targets was furthermore constrained by
comparing the degraded spectra to the spectra of O-type standards of 
solar metallicity compiled for the Tarantula Survey (Sana et al., in preparation) as well as to the spectra of VFTS targets obtained 
with Medusa-Giraffe which have already been classified.

Assigning a luminosity class to the ARGUS targets was more difficult because we could not exploit the selective emission effects in \ion{He}{ii}~$\lambda$4686 and \ion{N}{iii}~$\lambda\lambda$4634--4640--4642 (at subtypes earlier than O8), and the value of the \ion{He}{ii}~$\lambda$4686/\ion{He}{i}~$\lambda$4713 ratio (at subtypes O9-9.7). While at 
late-O types one could still rely on secondary criteria, such as the ratio of \ion{Si}{iv}~$\lambda$4089 to \ion{He}{i}~$\lambda$4026, no alternative luminosity diagnostics exist at early- and mid-O types. Given this situation, we decided to follow \citet{CA71} and exploit the increasing intensity  of \ion{Si}{iv}~$\lambda$4089 relative to the nearby \ion{He}{i}~$\lambda$4143 (at subtypes later than O5 only).

The classification of the non-variable ARGUS targets, derived as outlined above, is presented in Table \ref{rv_table}. In section~\ref{notes} of the appendix, we comment on specific sources, in particular those that have revised spectral types and the three that appear to have composite spectra. The accuracy of the spectral types reported in Table \ref{rv_table} for the ARGUS sources is typically between one and one and a half subtypes, with uncertainties caused by the effects of nebular emission, rotation \citep{Markova11}, and 
metallicity \citep{Markova10}. The uncertainty on the luminosity class is significantly larger as we were only able to separate the stars into two broad categories: high luminosity objects 
(luminosity class I/II) and low luminosity objects (luminosity class 
III/V). For completeness, we also include previously published spectral types in Table \ref{rv_table}.

\section{Supplementary Medusa data \label{medusa}}

To complement our RV measurements of the ARGUS sources, we also include the RV measurements of the non-variable O-type stars observed with Medusa-Giraffe (Paper VIII) in the inner 10\,pc of R136 . This gives us 22 additional  stars, 20 of which are located between 5 and 10\,pc from the centre of the cluster. We do not consider stars beyond 10\,pc from the centre in our analysis of R136. Although somewhat arbitrary, this cutoff at 10~pc is a reasonable trade-off between increasing the number of stars in our sample and limiting the possible contamination from nearby clusters or other star formation events in the surroundings of R136.

The Medusa observations (see Paper I) were performed using three of the standard Medusa-Giraffe settings (LR02: $\lambda$=3960--4564 \AA, $\Delta \lambda=0.61$~\AA, R$\sim$7\,000; LR03: $\lambda$=4499--5071~\AA, $\Delta \lambda=0.56$ \AA, R$\sim$8\,500; HR15N: $\lambda$=6442--6817 \AA, $\Delta \lambda=0.41$~\AA, R$\sim$16\,000). To detect RV variables, six epochs were observed for the LR02 setting with time constraints similar to those of the ARGUS observations (section \ref{argus_data} and Table \ref{epochs}).

The RV and variability analysis of these stars is presented in Paper VIII in the context of a study of the multiplicity of O-type stars in 30 Dor. The method is similar to the one we applied in section~\ref{var_crit}. As a consistency check, we applied our method on the LR02 Medusa spectra of several O-type stars and found absolute RVs and 1\,$\sigma$ uncertainties fully consistent with the results from Paper VIII. Thus, we do not repeat all the RV measurements of the Medusa stars, but adopt the values of Paper VIII instead. For two of those additional Medusa stars, the luminosity class could not be constrained, but none of the others is classified as a supergiant (Walborn et al., in preparation). The RVs of these stars, obtained from Gaussian fitting, should therefore be reliable as discussed in section \ref{abs}. The spectral types of the non-variable Medusa O-type stars, determined by  Walborn et al. (in preparation), are listed in Table~\ref{rv_table}.

Note that in parallel to the ARGUS observations, a few stars were also observed in the inner 10\,pc of R136 with the Ultraviolet and Visual Echelle Spectrograph (UVES), providing a greater resolving power than the Giraffe spectrograph. All the UVES targets but one in this region were also observed with ARGUS or Medusa, and these were already shown to be binaries/variable based on the ARGUS or Medusa observations alone. One star was observed only with UVES; Mk\,39, an O2.5~If*/WN6 star \citep{crowtherwalborn2011}, which is not suitable for our RV analysis because the \ion{He}{ii} lines are wind contaminated.

Our final sample of apparently single ARGUS and Medusa stars within 10\,pc from the centre of R136 is presented in Table~\ref{rv_table}. These are all the stars that we use for the analysis of the dynamics in the following sections. Their absolute RVs and their projected distance from the centre of the cluster (which we adopt to be the position R136-a1) are also listed. The projected distances are for an adopted distance modulus of 18.5~mag (see Paper I). The absolute RV adopted for a given star is the weighted mean RV of all epochs. The spatial distribution of variable and non-variable ARGUS sources is shown in Fig.~\ref{argus}, along with the positions of the non-variable Medusa O-type stars added to our sample.

\section{Velocity dispersion \label{vdisp}}

   \begin{figure*}[!t]
   \centering
\includegraphics[width=18cm]{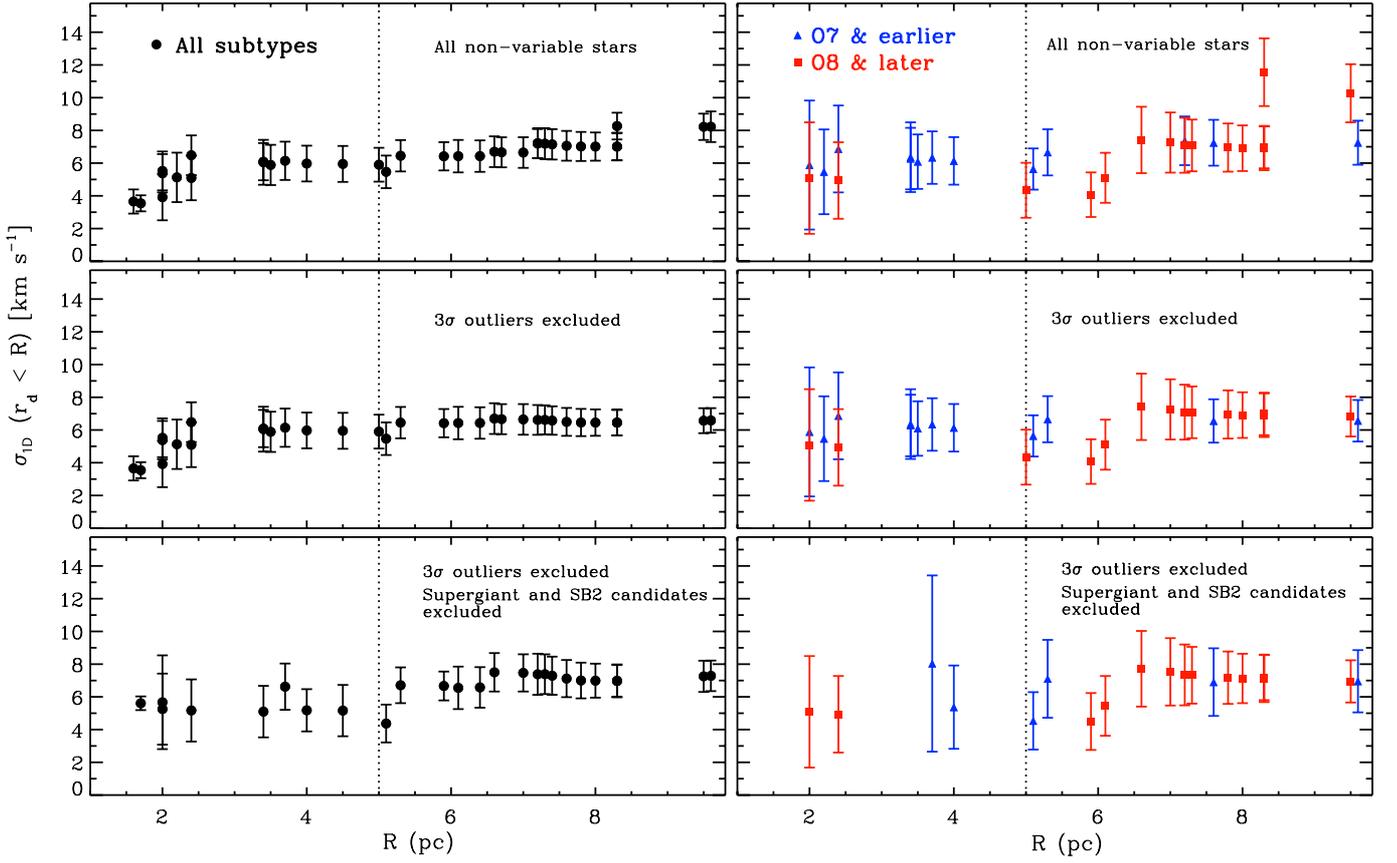}
      \caption{Observed line-of-sight velocity dispersion, as a function of R, for the stars within a projected radial distance R from the centre of R136. In the left panels, stars of all subtypes are included, while in the right panels, we divide them into two subsamples: O7 and earlier (blue triangles), O8 and later (red squares). In the top panels, all the non-variable stars (see Table~\ref{rv_table}) are included. In the middle panels, two stars with a radial velocity more than 3$\sigma$ away from the mean RV are excluded. In the bottom panel, supergiant and SB2 candidates are also excluded. In each panel (and subsample), the first point from the left is the velocity dispersion of the four stars (of that subsample) closest to the centre, the second point is the velocity dispersion of the innermost five stars, and so on.}
         \label{sigmav_plot}
   \end{figure*}

\subsection{Velocity dispersion upper limit \label{upper}}

We can now determine the observed line-of-sight velocity dispersion ($\sigma_{\rm 1D}$) from the RV measurements of the non-variable stars. Because any effect not yet taken into account (e.g. undetected binaries, intrinsic errors) would tend to increase the inferred $\sigma_{\rm 1D}$, we can consider that the results shown in this subsection represent an upper limit to the actual line-of-sight velocity dispersion of the cluster.

In what follows, $\sigma_{\rm 1D}$ is determined by computing the standard deviation, i.e. \citep{bevington}

\begin{equation}
\mu=\frac{1}{\sum_i 1/\sigma_i^2} \sum_i \frac{RV_{i}}{\sigma_i^2},
\label{mu}
\end{equation}

\begin{equation}
{\rm Var} = \frac{1}{\sum_i 1/\sigma_i^2}  \sum_i \frac{(RV_i- \mu)^2}{\sigma_i^2},
\label{var}
\end{equation}

\begin{equation}
\sigma_{\rm 1D} = \sqrt{{\rm Var} \ \frac{N}{N-1}},
\label{sig_def}
\end{equation}

\noindent{where $\sigma_i$ is the uncertainty on the RV measurement $RV_i$, $\mu$ is the weighted mean RV, Var is the variance, and $N$ is the number of measurements.

In the upper left panel of Fig.~\ref{sigmav_plot}, we present, as a function of projected radial distance from the centre of R136, the $\sigma_{\rm 1D}$ of all the non-variable stars within that radius. Apart from an apparent increase in $\sigma_{\rm 1D}$ in the inner region (most likely due to the low number of stars and associated large uncertainty on $\sigma_{\rm 1D}$ and its error), the velocity dispersion profile is relatively flat. For the stars within 5\,pc from the centre, we find $\sigma_{\rm 1D}\lesssim6$~$\kms$. Two stars dominate the increase in $\sigma_{\rm 1D}$ between 5 and 10\,pc: VFTS~536 (r$_{\rm d}$=7.2\,pc; RV=248.4$\pm$1.4~$\kms$) and VFTS~540 (r$_{\rm d}$=8.3\,pc; RV=242.7$\pm$1.1~$\kms$).

As a next step, to see the effect of possible outliers (slow runaways or massive stars along the line of sight but not members of R136), we exclude the stars with an RV more than 3\,$\sigma$ away from the (weighted) mean RV of our sample (267.7~$\kms$). We choose $\sigma$ to be 6\,$\kms$, the observed dispersion of the stars within 5\,pc from the centre. VFTS~536 and VFTS~540 are indeed $>$3$\sigma$ outliers, and excluding them results in an even flatter profile (Fig.~\ref{sigmav_plot}, middle left panel).

When also excluding possible supergiants (I/II or no luminosity class attributed in Table \ref{rv_table}), for which the RVs obtained from Gaussian fits could be problematic (see section~\ref{abs}), and SB2 candidates (composite spectra, see Table \ref{rv_table}), the results do not change significantly, although more fluctuations are seen in the profile and the error bars are larger due to the smaller number of stars (Fig.~\ref{sigmav_plot}, bottom left panel). The apparent increase in the inner 2--3\,pc also disappears when these supergiant and SB2 candidates are excluded.

In section~\ref{effect_binaries}, we estimate the contribution of errors and undetected binaries to $\sigma_{\rm 1D}$ and attempt to reproduce the observed velocity dispersion for the stars within 5\,pc (in projection) from the centre, i.e. $\sigma_{\rm 1D}=6$~$\kms$. We could choose a different radius, but because the velocity dispersion profile appears remarkably flat in the inner 10\,pc, this would not change the results significantly. There is also a natural cut at $\sim$5\,pc if we consider the definition of a cluster proposed by \citet{gielesPZ2011}, which states that stellar agglomerates for which the age of the stars exceeds the crossing time are bound and thus referred to as star clusters. The crossing time (i.e. the distance for a star to travel from one side of the cluster to the other; $2 r$) is roughly $\sigma_{\rm 1D} \ \times$~age. Given an age of $\sim$2\,Myr and $\sigma_{\rm 1D}\sim6\ \kms$, we can conclude that the stars that are physically within $\sim$6\,pc from the centre are part of the cluster following the above definition.

\subsection{Contamination by ``halo'' stars \label{halo}}

The light profile of R136 suggests that it is not a single-component cluster but the composite of a real cluster and a ``halo'', i.e. an OB association, with the latter contributing to more than 50\% and possibly as much as 90\% of its total integrated light \citep{maiz2001, 2003MNRAS.338...85M}. It is therefore worth asking how much could the OB association contaminate our velocity dispersion measurement for the cluster. In the double-component EFF fit \citep{1987ApJ...323...54E} to the light profile of R136 by \citet{2003MNRAS.338...85M}, the projected radius where the two components contribute equally is at about 5\,pc. If we consider the inner profile as the cluster and the outer profile as the halo, we can conclude from this fit that the contribution of halo stars projected on the cluster is negligible ($\lesssim5\%$) in the inner 1.25\,pc, but it could be significant beyond 5\,pc.

We might expect the OB association to have a low velocity dispersion, as is observed in Galactic cases once binaries and runaways are excluded \citep[e.g.][]{debruijne1999}. However, the latter are hard to identify if there is a massive cluster with a higher velocity dispersion embedded in the OB association. To explore the cluster/halo dichotomy, we divided our sample into two subtype groups such that they had roughly the same number of stars, which resulted in these two subsamples: earlier than O7 and later than O8. We then computed the velocity dispersion as in section~\ref{upper} for the stars that could be placed in one of these groups (Fig.~\ref{sigmav_plot}, right panels). We would expect the earlier type stars to be more concentrated towards the centre as a result of mass segregation or due to an overall age difference between the cluster and halo populations. Fig.~\ref{sigmav_plot} indeed suggest a higher concentration of early-types towards the core, but this could well be due to the increasing crowding effects in the innermost regions favoring the detection of the brightest (most probably earliest) stars. In any case, there is no obvious difference in velocity dispersion between the two subsamples, and the contribution of halo stars to the velocity dispersion remains very difficult to identify.

\subsection{The contribution of cluster rotation \label{rot}}

In an accompanying letter \citep{vhb2012a}, we present evidence for rotation of R136. From comparison of our RV measurements with different simple rotating models, we infer a rotational amplitude of $\sim$3\,$\kms$ and an optimal position angle for the rotation axis at an angle of $\sim$45$^{\circ}$ east of north. To remove the anisotropy due to the suggested rotation and its contribution to the computed velocity dispersion, we subtracted from our measured RVs the rotation curve from these simple models. We find that the velocity dispersion obtained after this correction is typically 0.5\,$\kms$ lower than the values presented in section~\ref{upper}. Thus, a small component of the observed line-of-sight velocity dispersion could be attributed to cluster rotation.

\subsection{The velocity dispersion when including binaries/variable stars \label{detect}}

It is interesting to see what the computed $\sigma_{\rm 1D}$ would have been if we had not been careful about identifying and rejecting binaries and variable stars, or if we were not dealing with multi-epoch observations.

We randomly selected one epoch for each ARGUS source on which the RV analysis was performed (variable and non-variable sources), and repeated the process for 10\,000 combinations of epochs. A median $\sigma_{\rm 1D}$ of 25.0\,$\kms$ was obtained from all these combinations of single-epoch RV measurements, with a standard deviation of 5.9\,$\kms$ and values of $\sigma_{\rm 1D}$ ranging from 12.9 to 48.0\,$\kms$. If we do a similar test but limit ourselves to the non-variable ARGUS sources (Table \ref{rv_table}), we find a median $\sigma_{\rm 1D}$ of 6.2\,$\kms$ (in good agreement with the results of section \ref{upper}), with a standard deviation of 0.7\,$\kms$.

Note however that the velocity dispersion obtained when including all the RV variables cannot be directly compared with the velocity dispersion one would obtain from the integrated light of a distant star cluster, as a few outliers could increase the velocity dispersion significantly without contributing much to the integrated light.

\subsection{The contribution of errors and undetected binaries \label{effect_binaries}}

To estimate the contribution of measurement errors and undetected binaries to the observed velocity dispersion, we performed a series of Monte Carlo simulations. These are adapted from the method presented by \citet{sana2009} and refined in Paper VIII to estimate the probability to detect binary systems. \citet{boschmeza2001} have also previously performed similar Monte Carlo simulations testing intrinsic and observed properties of binary stars. Our simulations mimic the process that we have been going through, i.e. identify variables from series of RV measurements and then compute the velocity dispersion from the remaining non-variable stars.

The general procedure goes as follows. We first adopt reasonable orbital parameter distributions (period, mass ratio, eccentricity) and an intrinsic binary fraction. Then, for 10\,000 populations of $N$ stars (where $N$ is the size of the sample, i.e. all the ARGUS sources plus Medusa O-type stars within 5\,pc from the centre of R136 on which the RV variability analysis was performed), we randomly draw which are binaries and which are single, and also randomly draw the parameters for the binaries from the adopted distributions. From these, we compute the orbital velocity at each epoch (based on the time sampling of the observations) for all the binaries assuming random orientations of the orbital planes and uncorrelated random time of periastron passage. We then add the measurement noise (based on the RV uncertainties of our sample) to the computed RVs of binaries and single stars. By applying the RV variability criteria of section~\ref{var_crit}, we can then eliminate the stars that we would have flagged as variable, compute the line-of-sight velocity dispersion of the apparently non-variable stars, and estimate the contribution from the orbital motion of binaries that were not detected. This procedure has the advantage that we not only take into account the long period binaries (i.e. too long a period to be detected with the VFTS), but also all the shorter binary systems missed by our time sampling, with a statistical weight exactly defined by the incompleteness.

Before performing the full procedure outlined above, we estimated the contribution of measurement errors by taking a population with no binaries. For each star, we drew the RV at each epoch from a Gaussian distribution centered on zero and a sigma corresponding to the RV error at the corresponding epoch. We then applied the RV variability criteria of section~\ref{var_crit} to make sure that our adopted thresholds do not lead to false detections (the false detection rate was indeed found to be negligible). We finally computed the velocity dispersion of the population and repeated this for 10\,000 populations. The resulting velocity dispersion distribution is represented by the blue dotted curve in Fig.~\ref{sigmav_dist}. The peak of this blue curve is at 0.80\,$(\kms)^{-1}$, out of the graph. It shows that, given the precision of our RV measurements, the intrinsic contribution of errors to the observed velocity dispersion is small. Note that we performed two tests on the population size: (i) we first used $N$ stars, the full population size, and (ii) we randomly picked stars from the initial population following the results of a binomial test with a success rate of 50\%, mimicking the fact that the dispersion is usually computed using an effective population of about half the original one because of binary rejection. The two approaches made no difference on the resulting contribution of measurement errors.

   \begin{figure}[!t]
   \centering
\includegraphics[width=9cm]{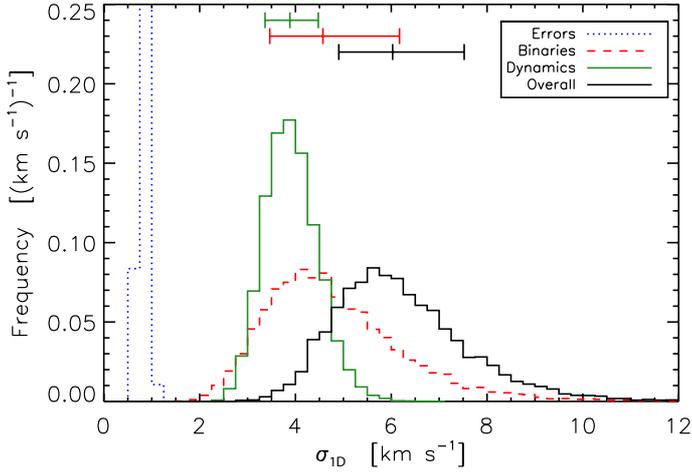}
      \caption{Estimate of the line-of-sight velocity dispersion distribution for massive stars in the inner 5\,pc of R136. The blue dotted curve shows the contribution of measurement errors. The red dashed curve is the distribution from binaries undetected after applying our variability criteria (the initial input population has a binary fraction of 100\%, and an \"{O}pik-law distribution of the periods is adopted with period range of 0.15$-$6.85 in $\log P$ with $P$ in units of days). The green dashed-dotted curve is the dynamical velocity dispersion which best reproduces the observed velocity dispersion. The black solid curve takes into account measurement errors, undetected binaries, and the dynamical velocity dispersion. The median (central tick) and 68\% confidence interval (equivalent to $\pm$1\,$\sigma$ for Gaussian distributions) of the distributions are indicated on the upper part of the graph.}
\label{sigmav_dist}
   \end{figure}

The contribution of undetected binaries depends on what is assumed for the input distributions. We focus here on the distribution of orbital periods, because the distributions of mass ratios and eccentricities have a limited impact (see Paper VIII). In what follows we adopt a flat distribution of mass ratios, as motivated by the recent results of \citet{sanaevans}, \citet{sanasci}, and \citet{kiminki}.

As a first test to estimate the intrinsic contribution of undetected binaries, we adopted a conservative binary fraction of 100\% and a standard \"{O}pik-law distribution for the period (i.e. flat distribution of $\log P$) with a period range of 0.15$-$6.85 (in $\log P$, where $P$ is in units of days). The maximum period adopted corresponds to the extrapolation of the \"{O}pik law until a 100\% binary fraction is reached when considering the observed binary fraction and the overall detection rate of VFTS (Paper VIII). We ran the full procedure outlined above, but without adding the measurement noise to the extracted orbital velocities. The resulting velocity dispersion distribution (from the stars that are not identified as RV variable) is shown with the red dashed curve in Fig.~\ref{sigmav_dist}. Under these assumptions, the velocity dispersion from undetected binaries is 4.6$_{-1.1}^{+1.6}$\,$\kms$. The quoted value is the median of 10\,000 populations, and uncertainties correspond to
the 68\% confidence interval (equivalent to $\pm$1\,$\sigma$ for Gaussian distributions). Note that within uncertainties, the undetected binaries alone can produce the observed velocity dispersion of 6\,$\kms$.

To recover the true velocity dispersion of the cluster, we repeated the simulation, this time adding the measurement noise to the orbital RVs. We also included a contribution from the dynamical velocity dispersion that was varied until the most probable velocity dispersion matched the observed velocity dispersion. For simplicity, we assumed that this dynamical contribution to the velocity dispersion did not vary as a function of radius. We estimate that the line-of-sight velocity dispersion attributable to the cluster dynamics is 3.9$_{-0.5}^{+0.6}$~$\kms$. The errors on this value correspond to changes in the input dynamical velocity dispersion that result in median values of the output velocity dispersion distribution at percentiles 0.16 and 0.84 of the optimal output distribution (i.e. the `overall' simulated distribution for which the median is the observed velocity dispersion). Note that we obtain a velocity dispersion of 26$\pm$9\,$\kms$ from these simulations for a single epoch when including all binaries (i.e. without applying our variability criteria and thus without rejecting the binaries that would be detected), in keeping with the value of $\sim$25\,$\kms$ that we found in section~\ref{detect} from the single-epoch RV measurements of all ARGUS sources (variable and non-variable).

We also ran simulations using different binary fractions and period distributions. We first considered only periods shorter than $10^{3.5}$~days, i.e. the ones that could be detected by VFTS (Paper VIII). Assuming a 50\% binary fraction and a standard \"{O}pik law for the period distribution, the estimated velocity dispersion from the undetected shorter period binaries alone is 2.4$^{+1.5}_{-1.0}$~$\kms$. If instead of the \"{O}pik law we adopt the distribution measured in Paper VIII for the VFTS O-type binaries ($f(\log{P})\propto(\log{P})^{-0.45}$), we obtain a velocity dispersion from the undetected shorter period binaries of 2.0$^{+1.4}_{-0.9}$~$\kms$. This is the minimal contribution of undetected binaries. These values (to be compared with the red dashed curve in Fig.\,\ref{sigmav_dist}) indicate that the contribution of undetected binaries is dominated by long period systems outside the sensitivity range of the VFTS.

Recall that the velocity dispersion from undetected binaries when including very long period systems as well (i.e. binary fraction 100\%, \"{O}pik-law distribution of periods ranging from 0.15 to 6.85 in $\log P$) is 4.6$_{-1.1}^{+1.6}$\,$\kms$. If we adopt the distribution with $f(\log{P})\propto(\log{P})^{-0.45}$ instead, the velocity dispersion from both short and long period undetected binary systems is 3.4$_{-1.0}^{+1.5}$\,$\kms$. In that case, the dynamical velocity dispersion that best reproduces the observed velocity dispersion would be 5\,$\kms$. The \"{O}pik law might therefore overestimate the contribution of binaries by $\sim$1~$\kms$ compared to the period distribution measured in VFTS O-type binaries (Paper VIII), but we should bear in mind that the extrapolation of the period distribution to long periods is very uncertain.

In summary, we estimate that the velocity dispersion due to cluster dynamics alone (i.e. removing the effect of binarity but still including a small contribution from rotation) is likely somewhere between 4 and 5\,$\kms$, with a contribution of $\sim0.5$\,$\kms$ from rotation (see section~\ref{rot}).

\section{Discussion \label{discuss}}

Now that we have our measurement of the velocity dispersion at hand we can test the hypothesis that the cluster is in virial equilibrium. This is often done by deriving a dynamical, or virial mass, from the velocity dispersion which is then compared to the photometrically determined mass.
To be able to derive the former we require knowledge about the mass distribution of the stars. For equilibrium models with an isotropic velocity dispersion the one-dimensional velocity dispersion, $\disp$, relates to the mass, $M$, and the viral radius, $\rv$, as $M=6\rv \disp^2/G$. The virial radius is defined as $\rv\equiv GM^2/(2W)$, with $W$ the potential energy of the system. This relation is often expressed in terms of the radius containing half the light in projection, or effective radius ($\reff$), as $M=\eta\reff\dispsq/G$, with $\eta\approx10$.  This is under the assumption that light traces mass, that the half-mass radius ($\rh$) in projection is $3/4$ times the 3D half-mass radius \citep{1987degc.book.....S} and that the ratio $\rv/\rh\approx5/4$. 
The first two assumptions are not valid when a cluster is mass segregated \citep{2006MNRAS.369.1392F} as the 2D {\it light} radius can be twice as small as the 3D {\it mass} radius \citep{2008MNRAS.391..190G, 2007MNRAS.379...93H}. For models with very flat density profiles, it is difficult to estimate the half-light radius. The surface brightness profiles of young clusters  are often approximated by cored templates with a power-law decline of the form $I(r)=I(0)(1+r^2/\rs^2)^{-\gamma/2}$, where $\rs$ is a scale radius. These profiles are often referred to as EFF profiles \citep{1987ApJ...323...54E}. For $\gamma>2$ these models contain a finite amount of light, but diverge to infinite luminosity when $\gamma\le2$. The boundary at $\gamma=2$ corresponds to 3D light profiles that decline as $r^{-3}$. For $\gamma$ larger than, but close to 2, the ratio $\rv/\rh$ becomes very sensitive to the exact value of $\gamma$
 \citep{2010ARA&A..48..431P}. Additionally, determining $\reff$ becomes difficult as this quantity, and the total luminosity, can become unrealistically large when extrapolating to infinity.
 
The light profile of R136 has a profile close to $\gamma=2$ \citep{2005ApJS..161..304M}, with indications for a `bump' in the optical light profile at about 10\,\pc\ \citep{1995ApJ...448..179H, maiz2001, 2003MNRAS.338...85M}. The presence of an additional component from the larger scale and near-constant-density OB association in which R136 is located (see section \ref{halo}) has been interpreted as the reason for this relatively flat profile. In the near-infrared (NIR), the profiles are even flatter than the critical value \citep{2009ApJ...707.1347A,2010MNRAS.405..421C} and the halo structure is not that obvious, but the NIR data did not extend very far into the OB association. Mass segregation, age differences between the core and halo, and differential extinction (which becomes important $\sim$10\,pc away from the centre of R136) could explain the flatter profile in the NIR compared to the optical.

With the above caveats in mind, it is still interesting to see what dynamical mass we obtain for R136. If we assume $\eta\approx10$, $3.4\lesssim\disp\lesssim6.0\,\kms$ (section~\ref{vdisp}), and adopt a half-light radius of $r_{\rm eff}=1.7$\,pc \citep{1995ApJ...448..179H} which is consistent with the half-light radius obtained for the inner component of the double EFF fit discussed in section~\ref{halo}, we get $M=4.6-14.2\times10^4\,\msun$. This is consistent with the estimated photometric mass of $\sim10^{5}$~$\msun$ by \citet{2009ApJ...707.1347A}, for which the cluster mass of $\sim5\times10^{4}$~$\msun$ (computed for stellar masses between 25\,$\msun$ and down to 2.1\,$\msun$) was extrapolated assuming a Salpeter slope down to 0.5\,$\msun$.

However, because of the difficulties outlined above, we decide to address whether R136 is in virial equilibrium by exploring an alternative method for which we do not need to know the total mass. This relies on estimating the central velocity dispersion by following a very similar method to that presented by \citet{richtremaine1986}. From the observations we find that the velocity dispersion is roughly constant with radius (section~\ref{upper}). We can express the expected central dispersion $\cdisp$ of (self-consistent and isotropic) models with isothermal inner parts in terms of observed properties of the cluster 
\begin{equation}
\cdispsq = \alpha \pi G \ml I(0)\rs,
\end{equation}
where $G\approx0.0043\,\pc\,\msun^{-1}\,\kmss$ is the gravitational constant, $\ml$ the mass-to-light ratio in $\msun/\lsun$ in the $V$-band and $\alpha$ depends on the model.

We first consider the modified Hubble profile \citep{1972ApJ...175..627R}, which is an EFF profile with $\gamma=2$, very close to the best fit to the surface brightness profile in the optical. From solving Jeans' equations assuming hydrostatic equilibrium and isotropic velocities we find that $\alpha=3-4\ln(2)\approx0.227$. Secondly, we look at the \citet{1911MNRAS..71..460P} model for which $\alpha=1/6\approx0.167$. Finally, we can consider the isothermal sphere, which cannot be expressed in an EFF profile. For this model $\rs$ is defined as\footnote{The radius $\rs$ in the isothermal model is the radius where the projected density falls by roughly half its central value. For the EFF model with $\gamma=2$ the projected density is  exactly half the central value at $\rs$ and this is why we use the same symbol.} $\rs^2=9\dispsq/[4\pi G\rho(0)]$, where $\rho(0)$ is the central density. For the isothermal sphere  $I(0)\approx 2\rho(0)\rs/\ml $ \citep{1987gady.book.....B} and  we thus have $\alpha\approx2/9\approx0.222$. In conclusion, $\alpha\approx0.2$  and the value is relatively insensitive to the choice of model (compared to $\eta$).

The central surface brightness in the $V$-band is about $I(0)\approx2.5\times10^6\,\lsun\,\pc^{-2}$ \citep{2005ApJS..161..304M} and together with $\rs\approx0.3\,\pc$ and $\ml\approx0.014$ \citep{2003MNRAS.338...85M,2005ApJS..161..304M} we find a predicted central velocity dispersion of
$\disp\approx 5.3\, \kms$. Our measured dispersion is consistent with this value and we conclude that R136 is in virial equilibrium in the inner 5\,pc. This is also consistent with a normal stellar initial mass function (IMF) for R136 \citep{2009ApJ...707.1347A}, as the expected velocity dispersion would be a factor of a few lower, for example, if the IMF was truncated at the low-mass end (the mass-to-light ratio would be lower in that case).

The expected velocity dispersion should actually be corrected for the fact that our stars are not at the centre of the cluster, unless we consider the cluster as an isothermal sphere in which case the dispersion is the same everywhere. For the Plummer model and the modified Hubble profile, the velocity dispersion for an isotropic model decreases with radius. For example, in the modified Hubble profile, the velocity dispersion at 5\,pc should be $\sim$40\% lower than at the centre for a core size of 0.3\,pc, so in virial equilibrium we would expect to measure a dispersion of $\sim$3\,$\kms$ at 5\,pc, which is still in relatively good agreement with our measured dispersion.

Other effects that we have not taken into account could also influence our estimate of the expected velocity dispersion in virial equilibrium. Mass segregation, for example, would result in a higher central surface brightness, a smaller core size, and a lower mass-to-light ratio. It is however not clear what would be the net effect on the velocity dispersion. Radial anisotropy would make the (projected) velocity dispersion profile decline more at larger radii with respect to an isotropic model with the same density profile \citep{clarkson2011, wilkinson2004}. On the other hand, tangential anisotropy could flatten the profile by reducing $\disp$ in the core and increasing it in the outer parts, which is an interesting perspective considering the evidence for rotation (H\'{e}nault-Brunet et al., submitted) and the relatively flat velocity dispersion profile that we obtained for R136. A similarly flat profile was also found for the Arches cluster albeit within a much smaller radial extent \citep{clarkson2011}. Detailed numerical modeling R136 is required for a meaningful quantitative discussion of the complicated effects outlined above, but this is beyond the scope of the present paper.

Other young massive clusters have recently been found to have a low velocity dispersion, suggesting that they are virial or even subvirial. Velocity dispersions of 4.5$\pm$0.8\,$\kms$ and 5.4$\pm$0.4\,$\kms$ were reported, respectively, for NGC 3603 \citep{rochau} and the Arches cluster \citep{clarkson2011} using proper motion measurements. From RV measurements of five yellow hypergiants and one luminous blue variable in Westerlund~1 showing little RV variations over 2 to 3 epochs, \citet{cottaar} estimated the velocity dispersion of this cluster to be 2.1$^{+3.3}_{-2.1}$\,$\kms$. From single-epoch near-infrared spectroscopy, \citet{mengel2007} found 5.8$\pm$2.1\,$\kms$ for the same cluster from the RVs of four red supergiants, 8.4\,$\kms$ from ten post-main-sequence stars \citep{mengel2008}, and finally 9.2\,$\kms$ from a sample of four red supergiants, five yellow hypergiants, and one B-type emission-line star \citep{mengel2009}. Note that these studies of Westerlund~1 use stars of spectral types and luminosity classes that are known to be pulsators or intrinsic RV variables \citep[e.g.][]{ritchie2009, clark2010}, which along with the small number statistics (both in terms of number of stars and number of epochs) might explain the range of values obtained for the velocity dispersion.

Given the young age of these clusters for which low velocity dispersions were found, including R136, this might look somewhat surprising if we expect the clusters to be expanding following gas expulsion. However, if the age of a cluster is at least few crossing times it might have had time to re-virialize, in which case the low velocity dispersions measured are not so surprising. A consequence of these results is that these young massive clusters are certainly not being disrupted by gas expulsion, and in fact appear to be stable from a very young age. From now on, gas expulsion will not have a large effect on the dynamical evolution of the cluster. A gas-free cluster in virial equilibrium is therefore a reasonable initial condition for dynamical simulations. The gas expulsion scenario would also predict strong radial orbits in the outer parts of the cluster, which would result in a steep decline in $\disp$ \citep{clarkson2011, wilkinson2004}}, and this is not observed. Several factors could potentially explain this apparent unimportance of gas expulsion in early cluster evolution, including a high star-formation efficiency \citep[e.g.][]{goodwinbastian:2006}, the formation of these clusters from the merging of dynamically cool (subvirial) substructures \citep[e.g.][]{Allison2009}, and/or a cluster formation process resulting in a de-coupled distribution of gas and stars that offsets the disruptive effect of gas expulsion \citep[e.g.][]{fellhauer2009, moeckel2011, kruijssen2012}. Another consequence of these results is that even if it were true that gas expulsion had a significant effect $\sim$1\,Myr ago, then R136 would have had to be incredibly dense in the past.

R136 is often considered as an extremely dense cluster, but it is interesting to compare the low velocity dispersion that we found to the much larger line-of-sight velocity dispersion of a globular cluster like 47 Tucanae \citep[11.6\,$\kms$;][]{mclaughlin2006}. R136 is still very young, and it will loose mass and expand due to stellar evolution. For an adiabatic mass loss, the radius will grow as $1/M$, so the velocity dispersion will go down by at least a factor of two ($\disp \propto (M/r)^{1/2} \propto M$), but in reality there is going to be even more significant expansion \citep{gieles_mr:2010} which will reduce the velocity dispersion even more by the time R136 is as old as 47 Tucanae.

\section{Conclusions \label{conc}}

In an effort to determine the dynamical state of the young massive cluster R136, we used multi-epoch spectroscopy of stars in the inner regions of 30 Dor. We measured RVs with a Gaussian fitting procedure on selected key helium lines and performed a quantitative assessment of the variability. Out of 41 sources for which spectra were extracted from the ARGUS IFU data cubes, 16 were identified as non-variable. All of these were classified as O-type stars, three of which were also revealed to have composite spectra. To this sample of 16 ARGUS sources, we added measurements from 22 apparently single stars observed in the surrounding regions with Medusa-Giraffe, also as part of VFTS.

Using this sample of 38 non-variable massive stars within 10\,pc from the centre of R136, we computed the velocity dispersion of the cluster. For the stars within 5\,pc, we place an upper limit of 6\,$\kms$ on the line-of-sight velocity dispersion. This result does not change significantly if we exclude the few 3\,$\sigma$ outliers, supergiant candidates, and stars having composite spectra. We also noted that the measured velocity dispersion of the cluster includes a small contribution of $\sim$0.5\,$\kms$ from rotation.

From Monte Carlo simulations, we established that the contribution of measurement errors to the observed velocity dispersion is almost negligible. We also estimated the contribution of undetected binaries, which is relatively small and dominated by long period systems beyond the detectability range of VFTS. When taking errors and undetected binaries into account, we estimate that the true velocity dispersion of the cluster (i.e. attributable only to cluster dynamics) is between 4 and 5\,$\kms$ for the stars within 5\,pc from the centre.

Under basic assumptions, the expected central velocity dispersion in virial equilibrium was found to be  $\approx$5.3\,$\kms$, in good agreement with our measurement, and we conclude that R136 is in virial equilibrium. Combined with the low velocity dispersions found in a few other young massive clusters, our results suggest that gas expulsion does not significantly alter the dynamics of these clusters.

We would have obtained a velocity dispersion of $\sim$25\,$\kms$ if binaries had not been identified and rejected, which supports the suggestion that the alleged super-virial state of young star clusters can be explained by the orbital motions of binary stars \citep{gieles2010}.

\begin{acknowledgements}
We are grateful to the referee, Guillermo Bosch, for constructive comments.
VHB acknowledges support from the Scottish Universities Physics Alliance (SUPA) and from the Natural Science and Engineering Research Council of Canada (NSERC). MG acknowledges financial support from the Royal Society. NB was supported by the DFG cluster of excellence `Origin and Structure of the Universe' (www.universe-cluster.de). JMA acknowledges support from [a] the Spanish Government Ministerio de Educaci\'on y Ciencia through grants AYA2010-15081 and AYA2010-17631 and [b] the Consejer{\'\i}a de Educaci{\'o}n of the Junta de Andaluc{\'\i}a through grant P08-TIC-4075. NM was supported by the Bulgarian NSF (DO 02-85). \end{acknowledgements}

\begin{appendix} 
\section{Individual epochs and summary of radial velocity/variability analysis }

\begin{table}[!h]
\centering                 
\caption{Individual epochs of the ARGUS observations.}            
\label{epochs}      
\begin{tabular}{l c c c}       
\hline\hline                 

Field & Epoch \# & Exposures & MJD\\    
\hline
A1	&	1 & 01[a-f] & 54761.237\\
	& 2	& 02 [a-f]	&54761.287 \\
	& 3	& 03 [g-l]	&54767.283\\
	& 4	& 03 [a-f]	&54845.177\\
	& 5	& 04 [a-f]	&54876.131\\
	& 6	& 05 [c+d] &	55173.310\\
	& 7	& 05 [a+b]	&55178.167\\
	\hline
A2	&1&	01 [a+b]	&54790.290\\
	&2	&02 [a+b]&	54791.140\\
	&3	&03 [a+b]	&54846.095\\
	&4	&04 [a+b]	&54889.102\\
	&5	&05 [a+b]	&55173.232\\	
\hline
A3	&1	&01 [a+b]&	54791.200\\	
	&2	&02 [a+b]&	54791.247\\	
	&3	&03 [a+b]	&54846.256\\	
	&4	&04 [a+b]	&54890.094\\	
	&5&	05 [a]	&55202.149\\	
\hline
A4	&1	&01 [a+b] &	54791.317\\	
	&2	&02 [a+b]	&54792.176\\	
	&3	&03 [a+b]	&54847.161\\	
	&4	&04 [c+d]	&54892.102\\	
	&5	&04 [e+f]	&54894.039\\	
	&6	&04 [g+h]	&54896.040\\	
	&7	&04 [a+b]	&54898.057\\	
	&8	&05 [a+b]	&55201.187\\
\hline	
A5	&1	&01 [a+b]	&54803.172\\
	&2	&02 [a+b]	&54803.216\\
	&3	&03 [a+b]	&54851.099\\
	&4	&04 [c+d]	&54893.045\\
	&5	&04 [a+b]	&54907.085\\
	&6	&05 [a+b]	&55204.144\\	
\hline
                                   
\end{tabular}
\tablefoot{The modified Julian date (MJD) represents the central time of all the exposures of a given epoch. The nomenclature of individual exposures follows that presented in Paper I.}
\end{table}

\clearpage

\onecolumn

\small{
\begin{longtable}{l c c | c  c | c c c c p{4cm}}    
\caption[]{Results of the variability tests for the ARGUS sources.
\label{var}} \\           
\hline \hline                 
 ID & Field  & Line & SB2 & Epoch(s) & $\left | {\frac{RV_{i} - \mu}{\sigma_{i}}} \right |_{\rm max}$  & $P(\chi^2, \nu)$ & $\Delta$RV$_{\rm max}$ & TVS   & Notes\\
 & &  &  &   &    &   & [$\kms$] &   &  \\
\hline\hline
\endfirsthead
\caption[]{{continued}} \\
\hline\hline
ID & Field & Line & SB2 & Epoch(s) & $\left | {\frac{RV_{i} - \mu}{\sigma_{i}}} \right |_{\rm max}$  & $P(\chi^2, \nu)$ & $\Delta$RV$_{\rm max}$ & TVS   & Notes\\
 & &  &   &   &    &   & [$\kms$] &   &  \\
\hline\hline
\endhead
\multicolumn{10}{r}{\it{continued on next page}}\\
\endfoot
\hline
\endlastfoot
542 &A5 &   \ion{He}{i}+\ion{He}{ii}~$\lambda$4026  & \dots  & \dots & 12.7  & 1.0000  &181.9$\pm$11.4 & \checkmark & TVS also shows significant variability in Balmer lines and \ion{N}{IV}~$\lambda$4058.\\
& &  \ion{He}{ii}~$\lambda$4200  & \dots  & \dots & 28.9  & 1.0000  & 185.9$\pm$5.5 & \checkmark &Double peak in TVS indicative of binary motion. Weak P~Cygni emission.\\
 &  & \ion{He}{ii}~$\lambda$4542  & \dots  & \dots &36.8  & 1.0000  & 148.5$\pm$4.2 & \checkmark &Double peak in TVS indicative of binary motion. Weak P~Cygni emission.  \\
\hline

545 &A3 &  \ion{He}{i}+\ion{He}{ii}~$\lambda$4026  & \dots  & \dots & 1.0  & 0.3484  & \dots & \dots & \\
&  &  \ion{He}{ii}~$\lambda$4200  & \dots  & \dots & 1.1  & 0.5688  & \dots & \dots & \\
&   & \ion{He}{ii}~$\lambda$4542  & \dots  & \dots &4.9  & 1.0000  & 14.4$\pm$3.0 & \dots & \\
\hline
570 &A5 & \dots& \dots & \dots   & \dots & \dots & \dots & \dots &  Too low S/N for RV analysis.\\
\hline

585&A5  &  \ion{He}{i}+\ion{He}{ii}~$\lambda$4026  & \dots  & \dots & 0.9  & 0.3268  & \dots & \dots & \\
& &  \ion{He}{ii}~$\lambda$4200  & \dots  & \dots & 2.7  & 0.9991  & \dots & \dots & \\
&  & \ion{He}{ii}~$\lambda$4542  & \dots  & \dots &2.5  & 0.9867  & \dots & \dots & \\
&  & 4200+4542  & \dots  & \dots &3.2  & 1.0000  & 32.7$\pm$9.1 & \dots & \\
\hline

1001&A2   & \dots& \dots & \dots   & \dots & \dots & \dots & \dots& Emission-line star, no suitable absorption line for RV analysis. Variability in \ion{He}{ii}~$\lambda$4542 emission from TVS.\\
\hline

\cellcolor[gray]{0.9}1002&A2 & \ion{He}{i}~$\lambda$4388  & \dots  & \dots & 2.4 & 0.9136 & \dots &  \dots& \\
\hline

1003&A2  & \dots& \dots & \dots   & \dots & \dots & \dots & \dots & B[e] star, no suitable absorption line for RV analysis.\\
\hline

\cellcolor[gray]{0.9}1004&A2 & \ion{He}{i}+\ion{He}{ii}~$\lambda$4026  & \dots  & \dots &  2.1 &  0.9600 & \dots &  \dots& \\
&  & \ion{He}{I}~$\lambda$4388  & \dots  & \dots & 2.3  &  0.9771 & \dots &  \dots& \\
&  & \ion{He}{ii}~$\lambda$4200  & \dots  & \dots & 2.4  & 0.9196  & \dots & \dots & \\
&  & \ion{He}{ii}~$\lambda$4542  & \dots  & \dots & 1.6  & 0.8421  & \dots & \dots & \\
&  & 4200+4388+4542  & \dots  & \dots & 3.3  &  0.9986 & \dots & & \\
\hline

1005 &A2 & \dots& \dots & \dots   & \dots & \dots & \dots & \dots &  Too low S/N for RV analysis.\\
\hline

\cellcolor[gray]{0.9}1006&A4 &  \ion{He}{i}+\ion{He}{ii}~$\lambda$4026 & \dots & \dots & 1.2 & 0.0869&\dots & \dots & \\
& &    \ion{He}{ii}~$\lambda$4200 & \dots & \dots & 1.9 & 0.6553& \dots &  \dots& \\
& &    \ion{He}{ii}~$\lambda$4542 & \dots & \dots & 3.7 & 0.9992& \dots &  \dots& \\
& &   4200+4542 & \dots & \dots & 3.5 & 0.9997& \dots & & \\
\hline

\cellcolor[gray]{0.9} 1007&A2 & \ion{He}{i}+\ion{He}{ii}~$\lambda$4026 & \dots & \dots & 1.9 & 0.9766 &\dots &  \dots& \\
 & &  \ion{He}{I}~$\lambda$4388 & \dots & \dots & 0.8 & 0.1244 &\dots &  \dots& \\
 & &  \ion{He}{ii}~$\lambda$4200 & \dots & \dots & 2.1 & 0.8986 &\dots & \dots & \\ 
 & &  \ion{He}{ii}~$\lambda$4542 & \dots & \dots & 0.8 & 0.1502 &\dots & \dots & \\ 
 & &  4200+4388+4542 & \dots & \dots & 1.4& 0.5402 &\dots & & \\ 
\hline

\cellcolor[gray]{0.9}1008&A2 &  \ion{He}{ii}~$\lambda$4200 & \dots & \dots & 1.0 & 0.1854 &\dots & \dots & \\ 
 & &  \ion{He}{ii}~$\lambda$4542 & \dots & \dots & 1.2 & 0.4761 &\dots & \dots & \\ 
 & &  4200+4542 & \dots & \dots & 1.6& 0.5707 &\dots & \dots & \\ 
 \hline                                   

\cellcolor[gray]{0.9} 1009&A2  &  \ion{He}{i}+\ion{He}{ii}~$\lambda$4026 & \dots & \dots & 0.7 & 0.1257&\dots &  \dots& \\ 
& &  \ion{He}{i}~$\lambda$4471 & \dots & \dots & 0.7 & 0.1607 &\dots &  \dots& \\ 
& &  \ion{He}{ii}~$\lambda$4200 & \dots & \dots & 1.6 & 0.4610 &\dots &  \dots& \\ 
& &  \ion{He}{ii}~$\lambda$4542 & \dots & \dots & 2.3 & 0.8584 &\dots &  \dots& \\ 
& & 4200+4542 & \dots & \dots & 2.1 & 0.7858 &\dots & \dots & \\ 
\hline

\cellcolor[gray]{0.9}1010&A4  &  \ion{He}{i}+\ion{He}{ii}~$\lambda$4026 & \dots & \dots & 1.2 & 0.3767 &\dots & \dots & \\ 
&  &  \ion{He}{ii}~$\lambda$4200 & \dots & \dots & 1.2 & 0.2252 &\dots & \dots & \\ 
& &  \ion{He}{ii}~$\lambda$4542 & \dots & \dots & 1.0 & 0.1728 &\dots & \dots & \\ 
&  & 4200+4542 & \dots & \dots & 1.5 & 0.4831 &\dots &  \dots& \\ 
\hline

1011&A4 & \dots& \dots & \dots   & \dots & \dots & \dots & \dots & Too low S/N for RV analysis.\\
\hline

\cellcolor[gray]{0.9}1012&A2 & \ion{He}{i}+\ion{He}{ii}~$\lambda$4026  & \dots  & \dots &  2.0 &  0.8506 & \dots & \dots & \\
&  & \ion{He}{i}~$\lambda$4388  & \dots  & \dots & 2.7  &  0.9949 & \dots & \dots & \\
&  & \ion{He}{i}~$\lambda$4471  & \dots  & \dots & 1.3  &  0.3517 & \dots &  \dots& \\
&  & \ion{He}{ii}~$\lambda$4200  & \dots  & \dots & 1.8  & 0.7524  & \dots & \dots & \\
&  & \ion{He}{ii}~$\lambda$4542  & \dots  & \dots & 1.1  & 0.2058  & \dots &  \dots& \\
&  & 4200+4388+4542  & \dots  & \dots & 2.9  &  0.9826 & \dots &  \dots& \\
\hline

1013&A4  & \dots& \dots & \dots   & \dots & \dots & \dots & \dots & Too low S/N for RV analysis.\\
\hline

\cellcolor[gray]{0.9}1014&A2 & \ion{He}{i}+\ion{He}{ii}~$\lambda$4026  & \dots  & \dots &  1.2 &  0.5620 & \dots & \dots & \\
&  & \ion{He}{i}~$\lambda$4388  & \dots  & \dots & 1.5  &  0.4265 & \dots &  \dots& \\
&  & \ion{He}{ii}~$\lambda$4200  & \dots  & \dots & 2.2  & 0.9645  & \dots &  \dots& \\
&  & \ion{He}{ii}~$\lambda$4542  & \dots  & \dots & 2.4  & 0.9268  & \dots &  \dots& \\
&  & 4200+4388+4542  & \dots  & \dots & 3.0  &  0.9926 & \dots & & \\
\hline

1015 &A2 & \ion{He}{i}+\ion{He}{ii}~$\lambda$4026  & \dots  & \dots &  1.1 &  0.2398 & \dots & \dots & \\
&  & \ion{He}{i}~$\lambda$4143  & \dots  & \dots & 1.1  &  0.1787 & \dots & \dots & \\
&  & \ion{He}{i}~$\lambda$4388  & \dots  & \dots & 1.9  &  0.8445 & \dots & \dots & \\
&  & \ion{He}{ii}~$\lambda$4200  & \dots  & \dots & 2.7  & 0.9953  & \dots & \dots & \\
&  & \ion{He}{ii}~$\lambda$4542  & \dots  & \dots & 3.8  & 1.0000  & 69.9$\pm$16.1 & \dots & \\
 &  &  4200+4542 & \dots  & \dots & 4.5 & 1.0000 & 77.7$\pm$13.9 & \dots & \\
\hline

1016&A4 & \ion{He}{i}+\ion{He}{ii}~$\lambda$4026  & ?  & 4?, 5? &  1.6 &  0.5778 & \dots & \dots & \\
&  & \ion{He}{ii}~$\lambda$4200  & \checkmark  & 2, 3, 6? & 12.9  & 1.0000  & 166.4$\pm$11.4 & \checkmark &  Double peak in TVS indicative of binary motion. \\
&  & \ion{He}{ii}~$\lambda$4542  & ?  & 6? & 19.2  & 1.0000 & 148.5$\pm$7.6 & \checkmark& Double peak in TVS indicative of binary motion.  \\
\hline

1017&A4 & \ion{He}{i}+\ion{He}{ii}~$\lambda$4026  & \dots  & \dots &  1.7 &  0.4677 & \dots & \dots & \\
&  & \ion{He}{ii}~$\lambda$4200  & \dots  &\dots & 3.7  & 1.0000  & 44.0$\pm$9.1 & \dots & Weak P~Cygni emission? \\
&  & \ion{He}{ii}~$\lambda$4542  & \dots  & \dots &6.2  & 1.0000 & 42.1$\pm$5.7 & \checkmark & Weak P~Cygni emission? \\ 
\hline

\cellcolor[gray]{0.9}1018&A4 &  \ion{He}{i}+\ion{He}{ii}~$\lambda$4026 & \dots & \dots & 1.8 & 0.7663&\dots & & \\
  &  &  \ion{He}{ii}~$\lambda$4200 & \dots & \dots & 2.1 & 0.5696& \dots & \dots & \\
  &  &  \ion{He}{ii}~$\lambda$4542 & \dots & \dots & 1.7 & 0.5315& \dots & \dots & \\
 &  &  4200+4542 & \dots & \dots & 1.9 & 0.8805& \dots & \dots& \\
\hline

1019 &A1 &  \ion{He}{ii}~$\lambda$4200 & \checkmark & 1-4, 6, 7 & 31.6 &1.0000  & 287.5$\pm$7.2 & \checkmark &  Double peak in TVS indicative of binary motion. \\
& & \ion{He}{ii}~$\lambda$4542 & \checkmark & 1-4, 6, 7 & 47.4 & 1.0000  & 301.5$\pm$4.8 & \checkmark &  Double peak in TVS indicative of binary motion. \\
\hline

\cellcolor[gray]{0.9}1020&A4 &  \ion{He}{i}+\ion{He}{ii}~$\lambda$4026 & \dots & \dots & 1.7 & 0.7254&\dots & \dots & \\
 &  &  \ion{He}{ii}~$\lambda$4200 & \dots & \dots & 1.8 & 0.4695& \dots &  \dots& \\
 &  &  \ion{He}{ii}~$\lambda$4542 & \dots & \dots & 1.3 & 0.2682& \dots &  \dots& \\
 &  &  4200+4542 & \dots & \dots & 1.4 & 0.5039& \dots & \dots & \\
\hline

1021&A4 &  \ion{He}{i}+\ion{He}{ii}~$\lambda$4026 & \dots & \dots & 2.8 & 0.9950&\dots & \dots & \\
 &  &   \ion{He}{i}~$\lambda$4471 & \dots & \dots & 3.1 & 0.9997&\dots & \dots & \\
 &   &  \ion{He}{ii}~$\lambda$4200 & \dots & \dots & 4.7 & 1.0000& 17.3$\pm$3.1 &  \dots& \\
 &   &  \ion{He}{ii}~$\lambda$4542 & \dots & \dots & 5.3 & 1.0000& 11.0$\pm$1.9 &  \dots& \\
\hline

1022 &A3, A4  &  \ion{He}{i}+\ion{He}{ii}~$\lambda$4026 & \dots & \dots & 3.7 & 0.9994&\dots & \dots & \\
&   &  \ion{He}{ii}~$\lambda$4200 & \dots & \dots & 3.4 & 0.9998& \dots &  \dots& \\
 &   &  \ion{He}{ii}~$\lambda$4542 & \dots & \dots & 3.6 & 1.0000& 14.2$\pm$3.2 &  \dots& \\
\hline

\cellcolor[gray]{0.9}1023&A3 &   \ion{He}{i}~$\lambda$4388 & \dots & \dots & 1.8 & 0.8739&\dots & \dots & \\
  &   &  \ion{He}{ii}~$\lambda$4200 & \dots & \dots & 1.2 & 0.5073& \dots &  \dots& \\
 &   &  \ion{He}{ii}~$\lambda$4542 & \dots & \dots & 0.5 & 0.0229& \dots &  \dots& \\
 &   &  4200+4388+4542 & \dots & \dots & 1.5 & 0.6155& \dots & \dots & \\
\hline

\cellcolor[gray]{0.9}1024&A1 &   \ion{He}{i}+\ion{He}{ii}~$\lambda$4026 & \dots & \dots & 3.1 & 0.9923&\dots & \dots & \\
&  &   \ion{He}{i}~$\lambda$4388 & \dots & \dots & 2.4 & 0.7910&\dots & \dots & \\
&   &  \ion{He}{ii}~$\lambda$4200 & \dots & \dots & 1.3 & 0.3178& \dots &  \dots& \\
&   &  \ion{He}{ii}~$\lambda$4542 & \dots & \dots & 2.9 & 0.9937& \dots &  \dots& \\
&   &  4200+4388+4542 & \dots & \dots & 2.2 & 0.9284& \dots & \dots & \\
\hline

1025  &A1 & \ion{He}{i}~$\lambda$4388  & \dots  & \dots & 1.3 & 0.4262 & \dots & \dots & Emission-line star. Significant variability in emission lines and P~Cygni profiles from TVS. \\
\hline

\cellcolor[gray]{0.9}1026&A3 &   \ion{He}{i}+\ion{He}{ii}~$\lambda$4026 & \dots & \dots & 1.5 & 0.4289&\dots & \dots & \\
&  &   \ion{He}{i}~$\lambda$4388 & \dots & \dots & 0.8 & 0.1612&\dots & \dots & \\
&  &  \ion{He}{ii}~$\lambda$4200 & \dots & \dots & 0.8 & 0.0743& \dots &  \dots& \\
&   &  \ion{He}{ii}~$\lambda$4542 & \dots & \dots & 2.2 & 0.9254& \dots &  \dots& \\
&   &  4200+4388+4542 & \dots & \dots & 1.9 & 0.9063& \dots & \dots & \\
\hline

1027&A5 &   \ion{He}{i}+\ion{He}{ii}~$\lambda$4026 & \dots & \dots & 1.7 & 0.7245 &\dots & \dots & \\
&   &  \ion{He}{ii}~$\lambda$4200 & \dots & \dots & 4.4 & 1.0000& 90.3$\pm$16.4 &  \dots& \\
&   &  \ion{He}{ii}~$\lambda$4542 & \dots & \dots & 5.3 & 1.0000& 74.7$\pm$12.4 &  \dots& \\
&   &  4200+4542 & \dots & \dots & 6.7 & 1.0000 & 75.2$\pm$9.9 &  \dots& \\
\hline

\cellcolor[gray]{0.9}1028&A3 &   \ion{He}{i}+\ion{He}{ii}~$\lambda$4026 & \dots & \dots & 1.8 & 0.7310&\dots & \dots & \\
&  &  \ion{He}{ii}~$\lambda$4200 & \dots & \dots & 1.2 & 0.2051& \dots &  \dots& \\
&   &  \ion{He}{ii}~$\lambda$4542 & \dots & \dots & 1.5 & 0.6237& \dots &  \dots& \\
&   &  4200+4542 & \dots & \dots & 1.3 & 0.3646& \dots & \dots & \\
\hline

1029&A5 &   \ion{He}{i}+\ion{He}{ii}~$\lambda$4026 & \dots & \dots & 13.0 & 1.0000& 113.1$\pm$7.1 & \checkmark & \\
&   &  \ion{He}{ii}~$\lambda$4200 & \dots & \dots & 18.0 & 1.0000& 97.5$\pm$4.1 & \checkmark &  Double peak in TVS indicative of binary motion. \\
&   &  \ion{He}{ii}~$\lambda$4542 & \dots & \dots & 30.6 & 1.0000& 102.5$\pm$2.5 &  \checkmark &  Double peak in TVS indicative of binary motion. \\
&   &  4200+4542 & \dots & \dots & 35.6 & 1.0000 & 101.0$\pm$2.1 & \dots & \\
\hline

1030&A5 & \dots& \dots & \dots   & \dots & \dots & \dots & \dots & Too low S/N for RV analysis.\\
\hline

1031&A5 &  \ion{He}{i}+\ion{He}{ii}~$\lambda$4026 & \checkmark & 1-3, 5-6 & \dots & \dots &\dots &\dots & SB3 at epochs \#1 and 2? \\
&   &  \ion{He}{ii}~$\lambda$4200 & \checkmark & 1-3, 5-6 & \dots & \dots& \dots & \checkmark &  SB3 at epochs \#1 and 2? \\
&   &  \ion{He}{ii}~$\lambda$4542 &  \checkmark & 1-3, 5-6 & 3.3 & 1.0000 & 35.6$\pm$10.6 &  \checkmark & SB3 at epochs \#1 and 2? \\
  \hline

1032&A5 &    \ion{He}{i}+\ion{He}{ii}~$\lambda$4026 & \dots & \dots & 3.8 & 0.9995&\dots & \dots & \\
&  &   \ion{He}{i}~$\lambda$4388 & \dots & \dots & 1.3 & 0.3758&\dots & \dots & \\
&   &  \ion{He}{ii}~$\lambda$4200 & \dots & \dots & 8.5 & 1.0000& 45.9$\pm$5.4 &  \dots& \\
&   &  \ion{He}{ii}~$\lambda$4542 & \dots & \dots & 11.8 & 1.0000& 39.3$\pm$3.4 &  \checkmark & \\
&   &  4200+4542 & \dots & \dots & 14.5 & 1.0000 & 40.7$\pm$2.9 & \dots & \\
\hline

1033&A5 &    \ion{He}{i}+\ion{He}{ii}~$\lambda$4026 &\checkmark& 1, 5 &  \dots &  \dots &\dots & \dots & \\
&  &   \ion{He}{i}~$\lambda$4143 &\checkmark & 1 &  \dots & \dots&\dots & \dots & \\
&  &   \ion{He}{i}~$\lambda$4388 &\checkmark & 1 & \dots&  \dots&\dots & \dots & \\
&   &  \ion{He}{ii}~$\lambda$4200 &\checkmark &1, 2 & \dots & \dots& \dots &  \dots& \\
&   &  \ion{He}{ii}~$\lambda$4542 &\checkmark &1, 2, 4 & 3.4 & 1.0000& 27.6$\pm$5.8 &  \dots & \\
\hline

1034&A5 & \ion{He}{i}+\ion{He}{ii}~$\lambda$4026  & \dots  & \dots &  4.7 &  1.0000 &13.3$\pm$2.1 & \dots & \\
&   & \ion{He}{i}~$\lambda$4143  & \dots  & \dots & 3.9  &  0.9999 & \dots &  \dots& \\
&   & \ion{He}{i}~$\lambda$4388  & \dots  & \dots & 3.7  &  0.9984 & \dots &  \dots& \\
&   & \ion{He}{i}~$\lambda$4471  & \dots  & \dots & 5.9  &  1.0000 & 7.5$\pm$1.6 &  \dots& \\
&   & \ion{He}{ii}~$\lambda$4200  & \dots  & \dots & 1.8  & 0.7640  & \dots & \dots & \\
&   & \ion{He}{ii}~$\lambda$4542  & \dots  & \dots & 3.3  & 0.9996  & \dots & \dots & \\
&   &  4200+4388+4542 & \dots & \dots & 4.9 & 1.0000 & 7.3$\pm$1.4 & \dots & \\
\hline
\cellcolor[gray]{0.9}1035&A5 &  \ion{He}{ii}~$\lambda$4200 & \dots & \dots & 1.1 & 0.2005& \dots &  \dots& \\
&   &  \ion{He}{ii}~$\lambda$4542 & \dots & \dots & 0.8 & 0.0504& \dots &  \dots& \\
&   &  4200+4542 & \dots & \dots & 1.3 & 0.2300& \dots & \dots & \\
\hline
1036&A5 & \dots& \dots & \dots   & \dots & \dots & \dots & \dots& Too low S/N for RV analysis.\\
\hline
1037&A5 & \dots& \dots & \dots   & \dots & \dots & \dots & \dots & Too low S/N for RV analysis.\\
\end{longtable}
\tablefoot{For SB2 systems, the values presented here refer to the primary component.}
}


\begin{center}
\scriptsize{
\begin{longtable}{l l l l l l l l l l}

\caption[]{RVs (in $\kms$) for individual epochs for all the ARGUS sources suitable for RV analysis.\label{RV_ind}} \\
\hline
ID & line &  RV$_{1}$ & RV$_{2}$ & RV$_{3}$& RV$_{4}$& RV$_{5}$& RV$_{6}$& RV$_{7}$& RV$_{8}$ \\
\hline
542 & \ion{He}{ii} 4542 & 254.1 $\pm$ 2.6 & 256.9 $\pm$ 3.4 & 158.4 $\pm$ 2.5 & 227.6 $\pm$ 3.5 & 229.7 $\pm$ 2.5 & 108.4 $\pm$ 2.4 & & \\
545 & \ion{He}{ii} 4542 & 230.5 $\pm$ 1.5 & 230.1 $\pm$ 1.8 & 228.4 $\pm$ 2.0 & 232.2 $\pm$ 2.1 & 242.8 $\pm$ 2.2 & & & \\
585 & \ion{He}{ii} 4200+\ion{He}{ii} 4542& 280.8 $\pm$ 6.6 & $\ldots$ & 248.1 $\pm$ 6.3 & 273.3 $\pm$ 4.5 & 273.7 $\pm$ 4.0 & 249.9 $\pm$ 5.3 & & \\
1002 & \ion{He}{i} 4388 & 257.6 $\pm$ 12.0 & 275.2 $\pm$ 13.6 & 264.9 $\pm$ 13.3 & 264.6 $\pm$ 11.8 & 302.1 $\pm$ 12.3 & & &\\
1004 &  \ion{He}{i} 4388+\ion{He}{ii} 4200+\ion{He}{ii} 4542& 285.7 $\pm$ 4.7 & 278.5 $\pm$ 5.2 & 276.2 $\pm$ 4.1 & 258.4 $\pm$ 4.8 & 273.4 $\pm$ 3.9& & &\\ 
1006	& \ion{He}{ii} 4200+\ion{He}{ii} 4542 & 249.5 $\pm$ 9.8 & 245.4 $\pm$ 11.3 & 294.4 $\pm$ 8.5 & 251.4 $\pm$ 13.6 & 275.0 $\pm$ 16.4 & 284.0 $\pm$ 17.9 & 284.3 $\pm$ 13.6 & 246.5 $\pm$ 8.3\\
1007 & \ion{He}{i} 4388+\ion{He}{ii} 4200+\ion{He}{ii} 4542 & 264.9 $\pm$ 2.6& 261.0 $\pm$ 2.8 & 262.9 $\pm$ 3.2 & 263.1 $\pm$ 2.9 & 268.1 $\pm$ 2.8 & & & \\
1008 &  \ion{He}{ii} 4200+\ion{He}{ii} 4542 & 277.3 $\pm$ 2.0 & 281.0 $\pm$ 2.4 & 275.0 $\pm$ 2.7 & 277.7 $\pm$ 2.1 & 275.8 $\pm$ 1.8 & & & \\	
1009 & \ion{He}{ii} 4200+\ion{He}{ii} 4542 & 277.6 $\pm$ 3.0 & 275.2 $\pm$ 3.3 & 266.9 $\pm$ 3.5 & 274.3 $\pm$ 3.7 & 274.9 $\pm$ 2.5 & & & \\
1010 & \ion{He}{ii} 4200+\ion{He}{ii} 4542 & 284.2 $\pm$ 5.0 & 287.1 $\pm$ 6.5 & 279.2 $\pm$ 5.3 & 273.7 $\pm$ 7.4 & 277.3 $\pm$ 7.5 & 278.7 $\pm$ 7.5 & 287.0 $\pm$ 6.4 & 292.2 $\pm$ 6.0 \\
1012 &  \ion{He}{i} 4388+\ion{He}{ii} 4200+\ion{He}{ii} 4542 & 275.0 $\pm$ 6.4 & 266.1 $\pm$ 6.4 & 267.2 $\pm$ 6.2 & 247.8 $\pm$ 5.8 & 269.9 $\pm$ 6.2 & & & \\
1014 & \ion{He}{i} 4388+\ion{He}{ii} 4200+\ion{He}{ii} 4542  & 269.3 $\pm$ 1.4 & 266.9 $\pm$ 1.7 & 268.1 $\pm$ 1.7 & 266.7 $\pm$ 1.4 & 262.3 $\pm$ 1.4 & & & \\ 
1015 & \ion{He}{ii} 4200+\ion{He}{ii} 4542 & 253.1 $\pm$ 7.9 & 279.7 $\pm$ 9.5 & 217.2 $\pm$ 9.6 & 294.9 $\pm$ 10.1 & 265.4 $\pm$ 9.0 & & & \\
1016 & \ion{He}{ii} 4542 & 316.9 $\pm$ 4.4 & 296.3 $\pm$ 5.4 & 329.0 $\pm$ 4.5 & 186.1 $\pm$ 6.1 & 342.6 $\pm$ 5.5 & 181.3 $\pm$ 6.2 & 329.8 $\pm$ 4.4 & 320.5 $\pm$ 4.1 \\
1017 & \ion{He}{ii} 4542 & 203.2 $\pm$ 3.3 & 213.3 $\pm$ 3.8 & 191.7 $\pm$ 3.6 & 217.7 $\pm$ 3.1 & 212.7 $\pm$ 3.6 & 227.2 $\pm$ 3.0 & 202.8 $\pm$ 2.9 & 185.1 $\pm$ 4.3 \\
1018 &  \ion{He}{ii} 4200+\ion{He}{ii} 4542 & 255.4 $\pm$ 1.6 & 260.9 $\pm$ 1.6 & 259.8 $\pm$ 1.5 & 259.0 $\pm$ 2.0 & 254.6 $\pm$ 2.0 & 258.2 $\pm$ 2.3 & 260.0 $\pm$ 1.7 & 258.1 $\pm$ 1.4 \\
1019 & \ion{He}{ii} 4542 & 362.2  $\pm$  5.6 &  381.9 $\pm$ 6.9 & 134.8 $\pm$ 4.0 & 117.8 $\pm$ 4.1 & 266.3 $\pm$ 3.8 & 86.5 $\pm$ 3.1 & 388.0 $\pm$ 3.6 & \\
1020 &  \ion{He}{ii} 4200+\ion{He}{ii} 4542 & 265.3 $\pm$ 2.1 & 266.8 $\pm$ 2.8 & 270.8 $\pm$ 2.5 & 266.4 $\pm$ 3.0 & 265.6 $\pm$ 3.1 & 263.7 $\pm$ 3.4 & 271.0 $\pm$ 2.7 & 268.0 $\pm$ 2.2 \\
1021 & \ion{He}{ii} 4542 & 248.6 $\pm$ 1.2 & 256.8 $\pm$ 1.3 & 249.8 $\pm$ 1.2 & 245.8 $\pm$ 1.4 & 247.2 $\pm$ 1.4 & 251.2 $\pm$ 1.5 & 250.0 $\pm$ 1.2 & 249.6 $\pm$ 1.1 \\ 
1022 &\ion{He}{ii} 4542  (A3 pointing)  & 222.2 $\pm$ 1.6 & 222.5 $\pm$ 1.6 & 227.0 $\pm$ 1.7 & 230.4 $\pm$ 1.8 & 226.0 $\pm$ 2.3 & & & \\
1022	 & \ion{He}{ii} 4542  (A4 pointing) & 219.6 $\pm$ 2.0 & 216.2 $\pm$ 2.6 & 229.6 $\pm$ 2.1 & 219.9 $\pm$ 2.5 & 227.6 $\pm$ 2.6 & 228.2 $\pm$ 2.5 & 221.5 $\pm$ 2.1 & 217.7 $\pm$ 2.2 \\
1023 & \ion{He}{i} 4388+\ion{He}{ii} 4200+\ion{He}{ii} 4542 & 270.2 $\pm$ 4.4 & 260.4 $\pm$ 3.9 & 267.4 $\pm$ 4.7 & 266.1 $\pm$ 4.5 & 272.0 $\pm$ 5.6 & & & \\
1024 & \ion{He}{i} 4388+\ion{He}{ii} 4200+\ion{He}{ii} 4542  & 272.3 $\pm$ 6.1 & 267.6 $\pm$ 6.1 & 245.5 $\pm$ 8.1 & 269.2 $\pm$ 8.1 & 272.6 $\pm$ 7.3 & 262.8 $\pm$ 4.7 & 255.6 $\pm$ 5.5 & \\ 
1026 &\ion{He}{i} 4388+\ion{He}{ii} 4200+\ion{He}{ii} 4542  & 264.9 $\pm$ 1.7 & 268.4 $\pm$ 1.8 & 265.7 $\pm$ 2.1 & 262.3 $\pm$ 2.1 & 260.8 $\pm$ 2.6 & & & \\ 
1027 & \ion{He}{ii} 4200+\ion{He}{ii} 4542  & 292.3 $\pm$ 4.2 & 293.3 $\pm$ 5.9 & 265.5 $\pm$ 5.2 & 218.1 $\pm$ 8.0 & 237.9 $\pm$ 4.2 & 260.4 $\pm$ 3.9 & & \\ 
1028 &  \ion{He}{ii} 4200+\ion{He}{ii} 4542& 272.1 $\pm$ 0.9 & 272.7 $\pm$ 1.0 & 271.6 $\pm$ 1.1 & 270.3 $\pm$ 1.2 & 272.1 $\pm$ 1.4 & & & \\ 
1029 & \ion{He}{ii} 4200+\ion{He}{ii} 4542& 290.7 $\pm$ 1.3 & 292.7 $\pm$ 1.2 & 227.0 $\pm$ 1.7 & 287.7 $\pm$ 2.9 & 262.5 $\pm$ 1.7 & 328.0 $\pm$ 1.3 & & \\ 
1031 & \ion{He}{ii} 4542 & 274.7 $\pm$ 8.0 & 278.0 $\pm$ 7.0 & 242.4 $\pm$ 8.1 & 256.5 $\pm$ 3.2 & 251.6 $\pm$ 5.3 & 275.4 $\pm$ 4.0 & & \\
1032 & \ion{He}{ii} 4200+\ion{He}{ii} 4542& 263.9 $\pm$ 1.7 & 264.2 $\pm$ 1.8 & 272.7 $\pm$ 2.7 & 285.6 $\pm$ 3.3 & 259.9 $\pm$ 2.2 & 300.6 $\pm$ 1.9 & & \\ 
1033 & \ion{He}{ii} 4542 & 268.3 $\pm$ 3.8 & 266.2 $\pm$ 3.8 & 240.7 $\pm$ 4.4 & 248.8 $\pm$ 7.9 & 251.2 $\pm$ 4.3 & 251.2 $\pm$ 3.3 & & \\ 
1034	 & \ion{He}{i} 4388+\ion{He}{ii} 4200+\ion{He}{ii} 4542 & 263.7 $\pm$ 1.0 & 263.2 $\pm$ 1.0 & 263.8 $\pm$ 1.2 & 270.3 $\pm$ 1.8 & 264.4 $\pm$ 1.2 & 270.5 $\pm$ 1.0 &  & \\ 
1035 &  \ion{He}{ii} 4200+\ion{He}{ii} 4542 & 268.1 $\pm$ 4.1 & 262.7 $\pm$ 4.6 & 272.1 $\pm$ 6.5 & 271.3 $\pm$ 9.2 & 271.9 $\pm$ 4.8 & 269.7 $\pm$ 5.8 & & \\
\hline
\end{longtable}
\tablefoot{The lines used for the final RV measurements of a given star are indicated. RVs for the Medusa targets are presented in Paper VIII.}
}
\end{center}

\twocolumn

\section{Notes on individual ARGUS sources \label{notes}}

   \begin{figure*}[!t]
   \centering
\includegraphics[width=15cm]{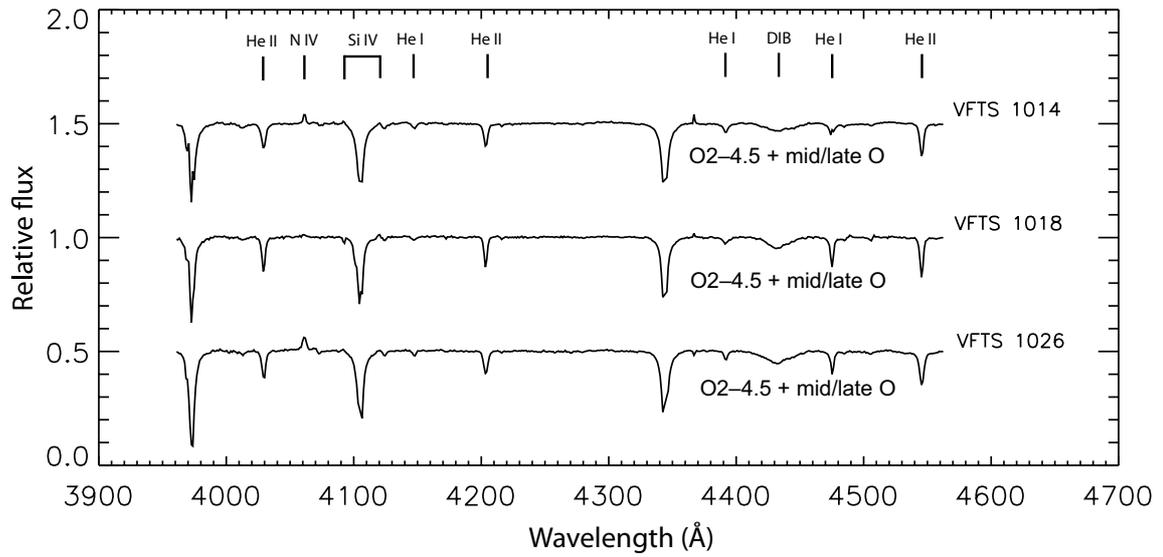}
      \caption{Non-variable ARGUS sources displaying composite spectra. Emission lines labelled are \ion{N}{iv}~$\lambda$4058 and \ion{Si}{iv}~$\lambda\lambda$4089, 4116. Absorption lines labelled are \ion{He}{i}~$\lambda$4026, \ion{He}{i}~$\lambda$4143, \ion{He}{ii}~$\lambda$4200, \ion{He}{i}~$\lambda$4388, \ion{He}{i}~$\lambda$4471, \ion{He}{ii}~$\lambda$4542 and a diffuse interstellar band at 4428 \AA.}
         \label{compo}
   \end{figure*}

We comment here on selected individual sources, paying particular attention to sources with previous identifications, composite spectra, and also to those which appear multiple in the WFC3 image.

\begin{itemize}
\item {\it VFTS 542:} This star is identified as a definite variable with a large amplitude from both ARGUS and Medusa observations. Its \ion{He}{ii}~$\lambda$4200 and \ion{He}{ii}~$\lambda$4542 lines show a weak P~Cygni component and it is classified as O2 If*/WN5 (Paper I), so even if it was not variable, its absolute RV could not be trusted. RV discrepancies as large as $\sim$40~$\kms$ are found at some epochs between \ion{He}{ii}~$\lambda$4200 and \ion{He}{ii}~$\lambda$4542.

\item {\it VFTS 545:} It is also known as Mk35 \citep{Mk85} and classified as O2 If*/WN5 (Paper I). There is a discrepancy of $\sim$20~$\kms$ between \ion{He}{ii}~$\lambda$4200 and \ion{He}{ii}~$\lambda$4542, and its absolute RV also cannot be trusted. Low-amplitude variability is only detected in \ion{He}{ii}~$\lambda$4542, which is stronger and has smaller RV uncertainties compared to \ion{He}{ii}~$\lambda$4200.

\item {\it VFTS 570:} The ARGUS spectra of this source were not analysed for RV variability because their S/N was too low, but from the Medusa spectra it was found to be a definite RV variable with a large amplitude (Paper VIII). Two stars appear to contribute significantly to the ARGUS source when comparing with the WFC3 image.

\item {\it VFTS 585:} This source was also found to be variable from the Medusa spectra (Paper VIII). Significant RV variability was detected from the relatively low S/N ARGUS spectra only once \ion{He}{ii}~$\lambda$4200 and \ion{He}{ii}~$\lambda$4542 were fitted simultaneously.

\item {\it VFTS 1001:} This source corresponds to a known Wolf-Rayet star, R134 \citep{feast60}, classified as WN6(h) \citep[e.g.][]{massey1998}. Although it was detected as an X-ray source and suggested as a possible colliding-wind binary by \citet{pzpooley}, it is not known to be a binary. Interestingly, our TVS analysis reveals significant variability in the \ion{He}{ii}~$\lambda$4542 emission, but it is unclear if this is due to a normalisation problem in a spectral range dominated by several emission lines, where the continuum is harder to define.

\item {\it VFTS 1003:}  This source was found to be a new B[e]-type star in Paper I. The TVS analysis performed in the present work did not reveal any significant variability other than in the nebular emission lines (due to sky subtraction).

\item {\it VFTS 1004:} The WFC3 image suggests that two sources are contributing to VFTS 1004, but it does not display a composite spectrum, it is not found to be variable, and the \ion{He}{i}~$\lambda$4388, \ion{He}{ii}~$\lambda$4200 and \ion{He}{ii}~$\lambda$4542 lines all have consistent absolute RVs.

\item {\it VFTS 1007:} Similarly to VFTS 1004, VFTS 1007 appears multiple when inspecting the WFC3 image, but it is not variable, it does not have a composite spectrum, and \ion{He}{i}~$\lambda$4388, \ion{He}{ii}~$\lambda$4200 and \ion{He}{ii}~$\lambda$4542 all have consistent absolute RVs.

\item {\it VFTS 1014:} The presence of \ion{N}{iv}~$\lambda$4058 and \ion{Si}{iv}~$\lambda$4089/4116 emission together with weak but well developed \ion{He}{i} singlet lines at 4121, 4143 and 4388 \AA\ suggests a composite spectrum (see Fig.~\ref{compo}). Based on the helium line diagnostics and the absence of \ion{Si}{iii}~$\lambda$4552, the later component is identified as a mid/late O-type star. From the relative strength of \ion{N}{iv} and \ion{Si}{iv}, the other component is O2-4.5 (we cannot be more precise because our the ARGUS spectrum does not include the \ion{N}{v} absorption region), in agreement with the O3~V classification of \citet{massey1998}. Even though its spectrum appears composite, this source did not show significant variability. The \ion{He}{i}~$\lambda$4388, \ion{He}{ii}~$\lambda$4200 and \ion{He}{ii}~$\lambda$4542 lines all have consistent absolute RVs.

\item {\it VFTS 1015:} This source is clearly multiple by comparison with the WFC3 image and significant RV variability is found in \ion{He}{ii}~$\lambda$4542. For some epochs, the RV of the \ion{He}{i}~$\lambda$4388 line is clearly different from that of  \ion{He}{ii}~$\lambda$4200 and \ion{He}{ii}~$\lambda$4542 lines.

\item {\it VFTS 1017:} This source is variable. Its \ion{He}{ii}~$\lambda$4200 and \ion{He}{ii}~$\lambda$4542 lines have a weak P~Cygni component. A discrepancy of up to $\sim$30~$\kms$ is found in the RVs of \ion{He}{ii}~$\lambda$4200 and \ion{He}{ii}~$\lambda$4542 at different epochs.

\item {\it VFTS 1018:} The presence of weak \ion{N}{iv}~$\lambda$4058 and \ion{Si}{iv}~$\lambda$4116 emission in combination with weak but well developed \ion{He}{i} singlet lines suggests a composite spectrum (see Fig.~\ref{compo}). Based on the helium line diagnostics and the absence of \ion{Si}{iii}~$\lambda$4552, the later component is identified as a mid/late O-type star.. From the relative strength of the \ion{N}{iv} and \ion{Si}{iv} emission, the other component is classified as O2-4.5, in relatively good agreement with the O3 III(f*) classification of \citet{massey1998}. Even though its spectrum appears composite, this source did not show significant variability. The \ion{He}{i}~$\lambda$4388, \ion{He}{ii}~$\lambda$4200 and \ion{He}{ii}~$\lambda$4542 lines all have consistent absolute RVs.

\item {\it VFTS 1019:} This is a known high-mass binary (R136-038) classified as O3 III(f*)~+~O8 by \citet{massey1998}, then revised as O3 V + O 6 V by \citet{MPV2002}. The ARGUS spectra show obvious variability, a large RV amplitude, and SB2 profiles at several epochs.

\item{\it VFTS 1022:} This source corresponds to Mk37a=R136-014 \citep{Mk85, massey1998}, classified as O4~If+ by \citet{massey1998}, but suggested as O3.5~If*/WN7 by \citet{crowtherwalborn2011}. 13 epochs (the source is on the edge of the A3 and A4 ARGUS pointings) made it possible to detect low-amplitude RV variability in \ion{He}{ii}~$\lambda$4542. However, even if it had not been flagged as variable, this star would not have been suitable for our analysis of the dynamics. A discrepancy of $\sim$15~$\kms$ is found between the RVs of \ion{He}{ii}~$\lambda$4200 and \ion{He}{ii}~$\lambda$4542, and the RV of \ion{He}{i}~$\lambda$4388 is significantly larger than that of the \ion{He}{ii} lines.

\item {\it VFTS 1023:} We classified this star as O8 III/V. \citet{massey1998} classified it as O6, but at this subtype \ion{He}{i+ii}~$\lambda$4026 should be as deep as \ion{He}{ii}~$\lambda$4200 while \ion{He}{i}~$\lambda$4471 should be significantly weaker than \ion{He}{ii}~$\lambda$4542. Also, \ion{He}{i}~$\lambda$4143 and \ion{He}{i}~$\lambda$4388 should be much weaker than \ion{He}{ii}~$\lambda$4200 and \ion{He}{ii}~$\lambda$4542 respectively, which is not what we see. A possible explanation for the discrepancy between our classification and that of \citet{massey1998} is that this source is an undetected single-lined spectroscopic binary.

\item {\it VFTS 1025:} This source appears multiple and the centre of the ARGUS position is offset between two stars in the WFC3 image, with the much brighter star being R136c. It is interesting to note that we find significant variability in the TVS of this source (see Fig. \ref{TVS}). R136c was identified as a probable binary \citep{Schnurr2009} and suspected to be a colliding-wind massive binary \citep{Crowther2010}.

\item {\it VFTS 1026:} When comparing with the WFC3 image, the centre of the ARGUS source appears offset between two stars. One of these is MH41, O3 III(f*) \citep{massey1998}, also classified as O8: V by \citet{walborn1997}. The light is probably dominated by MH41 (the brighter of the two stars), although we flagged VFTS 1026 as having a composite spectrum (see Fig.~\ref{compo}), as suggested by the presence of \ion{N}{iv}~$\lambda$4058 and \ion{Si}{iv}~$\lambda$4089/4116 emission together with weak but well developed \ion{He}{i} singlet lines at 4121, 4143 and 4388 \AA. Based on the helium line diagnostics and the absence of \ion{Si}{iii}~$\lambda$4552, the later component is identified as a mid/late O-type star. From the relative strength of the \ion{N}{iv} and \ion{Si}{iv} emission, the other component is classified as O2-4.5, in agreement with the classification of \citet{massey1998}. This source is however not variable, and its \ion{He}{i}~$\lambda$4388, \ion{He}{ii}~$\lambda$4200 and \ion{He}{ii}~$\lambda$4542 lines have consistent absolute RVs.

\item {\it VFTS 1031:} This corresponds to R136-025 \citep[O3 V;][]{massey1998}, which was flagged as a suspected variable by \citet{MPV2002}. In our ARGUS spectra, the \ion{He}{i}+\ion{He}{ii}~$\lambda$4026, \ion{He}{ii}~$\lambda$4200 and \ion{He}{ii}~$\lambda$4542 lines seem to display three components at some epochs.

\item{\it VFTS 1034:} This corresponds to Mk32, which is itself a blend of R136-013 (O8 III(f), \citealt{massey1998}; O7.5 II, \citealt{walborn1997}) and R136-074 \citep[O6 V,][]{massey1998}. The variability in this source is more obvious in the \ion{He}{i}+\ion{He}{ii}~$\lambda$4026 and \ion{He}{i}~$\lambda$4471 lines, but also significant when \ion{He}{i}~$\lambda$4388, \ion{He}{ii}~$\lambda$4200 and \ion{He}{ii}~$\lambda$4542 are fitted simultaneously.

\end{itemize}

\end{appendix}

\end{document}